\documentclass[10pt]{iopart}
\pdfoutput=1
\usepackage{graphicx}
\usepackage{xcolor}
\usepackage{iopams} 

\eqnobysec
\newcommand{\beq}{\begin{equation}}
\newcommand{\beqa}{\begin{eqnarray}}
\newcommand{\eeq}{\end{equation}}
\newcommand{\eeqa}{\end{eqnarray}}

\renewcommand{\d}{{\rm d }}

\newcommand{\prob}{\mathop{\rm Prob}\nolimits}

\renewcommand{\prob}{{\rm Prob}}

\newcommand{\s}{{\sigma}}

\newcommand{\un}{{+-}}

\newcommand{\en}{{e}}
\newcommand{\tcluster}{{t_{c}}}

\newcommand{\zeq}[1]{\mathrel{\mathop{=}\limits_{#1}^{}}}

\begin{document}
\date{\today}

\title[Dynamics of the two-dimensional directed Ising model]{Dynamics of the two-dimensional directed Ising model: zero-temperature coarsening}
\date{\today}
\author{C Godr\`eche$^1$ and M Pleimling$^2$}
\address{
$^1$Institut de Physique Th\'eorique, Universit\'e Paris-Saclay, CEA and CNRS\\
91191 Gif-sur-Yvette, France}\smallskip
\address{$^2$Department of Physics,
Virginia Polytechnic Institute and State University
Blacksburg, Virginia 24061-0435, USA
}
\begin{abstract}
We investigate the laws of coarsening of a two-dimensional system of Ising spins evolving under single-spin-flip irreversible dynamics at low temperature from a disordered initial condition.
The irreversibility of the dynamics comes from the directedness, or asymmetry, of the influence of the neighbours on the flipping spin.
We show that the main characteristics of phase ordering at low temperature, such as self-similarity of the patterns formed by the growing domains, and the related scaling laws obeyed by the observables of interest, which hold for reversible dynamics, are still present when the dynamics is directed and irreversible, but with different scaling behaviour.
In particular the growth of domains, instead of being diffusive as is the case when dynamics is reversible, becomes ballistic.
Likewise, the autocorrelation function and the persistence probability (the probability that a given spin keeps its sign up to time $t$) have still power-law decays but with different exponents.
\end{abstract}

\vspace*{1pt}\address{\today}

\maketitle

\section{Introduction}

The aim of this paper is to understand the laws of coarsening of a two-dimensional system of Ising spins on a square lattice evolving under single-spin-flip irreversible dynamics at low temperature from a disordered initial condition.
The irreversibility of the dynamics comes from the directedness, or asymmetry, of the influence of the neighbours on the flipping spin.
Yet, at any finite temperature and at long times, the system reaches a stationary state whose measure is that of the equilibrium two-dimensional Ising model.
In other words, the weight of configurations is given by the Boltzmann-Gibbs distribution associated to the ferromagnetic Hamiltonian (or energy) of the Ising model on the square lattice (see~(\ref{eq:energy})).

We will show that the main characteristics of phase ordering at low temperature, such as self-similarity of the patterns formed by the growing domains, and the related scaling laws obeyed by the observables of interest, which hold for reversible dynamics~\cite{bray,reviewgl,malte}, are still present when the dynamics is directed and irreversible~\cite{gb2009,cg2011,cg2013,gp2014,gl2015}, but with different scaling behaviour.
In particular the growth of domains, instead of being diffusive as is the case when dynamics is reversible, becomes ballistic.
Likewise, the autocorrelation function~\cite{autocorr,bray,reviewgl,malte} and the persistence probability~\cite{dbg,bdg,review} (the probability that a given spin keeps its sign up to time $t$) have still power-law decays but with different exponents.

A first hint of the occurrence of these phenomena is obtained by measuring the speed at which a given minority cluster (a circle, a right triangle, or a square) of ($+$) spins in a sea of ($-$) spins disappears.
We thus find that, while the time required for the disappearance of this cluster scales as its area for reversible dynamics, this time scales as the square root of the area for irreversible dynamics.
In other words, in this very simple situation, dynamics crosses over from diffusive to ballistic, under asymmetric irreversible dynamics.

We then demonstrate that this also holds for the general situation of a system relaxing at zero temperature from an initial disordered configuration.
To this end we study two complementary facets of the dynamics, namely the relaxation of the energy on one hand, and the relaxation of the equal-time correlation function on the other hand.

We first investigate the relaxation of the energy, which yields a first definition of the typical growing length scale in the system,
denoted by $\xi_{E}(t)$ (see (\ref{eq:xiEdef})).
The relaxation of the energy is characterized by two regimes: the scaling regime, where coarsening takes place, followed by the late-time regime, where finite-size effects are important. 
In the first regime, $\xi_{E}(t)$ is observed to grow diffusively ($\xi_{E}(t)\sim t^{1/2}$) for reversible dynamics, and ballistically ($\xi_{E}(t)\sim t$) as soon as dynamics is irreversible.
In the late-time regime the system either reaches the ground state or a blocked metastable state.
Indeed, though at long times a finite system is expected to reach one of the two ground states where spins are either all up, or all down,
this is true only with a finite probability.
As studied in~\cite{kr1,kr2,kr3,kr4,picco1}, if one draws a great many Ising spin systems on a two-dimensional square lattice, quenched from high temperature down to zero temperature and let them evolve under Glauber dynamics~\cite{glau}, then a fraction of them gets trapped in blocked configurations, which are stripes running across the system.
A similar situation holds for our model.
A quantitative study of this phenomenon is postponed to another work.
Here we determine the scale of time necessary for the energy to reach its ground state value.
We also measure the scale of time necessary for the system to reach a blocked state.
This allows us to confirm the change of regime in the coarsening behaviour of the system as soon as dynamics is irreversible:
coarsening becomes ballistic instead of being diffusive as is the case for reversible dynamics.

We then investigate the behaviour of the equal-time correlation function $C(r,t)$.
This gives access to an alternate definition of a growing length for a system undergoing phase ordering, denoted by $\xi_{C}(t)$ (see section~(\ref{sec:defLt})).
The ballistic regime is confirmed for irreversible dynamics, with $\xi_{C}(t)\sim t$.
These two facets of the dynamics are coherently related since $\xi_{E}(t)$ and $\xi_{C}(t)$ are essentially proportional.
One additional finding is the presence of a dynamical anisotropy in the growing length scale $\xi_{C}(t)$ for any value of the irreversibility parameter $V$ in~(\ref{eq:rate}).
This anisotropy 
manifests itself by a discrepancy between the values of this length according to the direction in which it is measured on the two-dimensional lattice.

We finally proceed to the study of the autocorrelation function and of the probability of persistence. 
Whether power-laws should survive or not in the ballistic regime is not obvious to predict a priori.
It turns out that both quantities exhibit power-law decays,
with autocorrelation and persistence exponents larger than their counterparts for reversible dynamics.
We also study the statistics of the mean temporal magnetization which gives another viewpoint on persistence and confirms the values of the exponents found by the decay of the persistence probability.
The very existence of power-law behaviours is a confirmation of the existence of self-similar coarsening, even in the presence of ballistic irreversible dynamics.

\section{Definition of the dynamical rules of the directed Ising model}
\label{sec:def}

In this section we first give the general definition of the dynamical rules of our model, valid for any temperature, we then specialize this definition to the zero-temperature situation under study.

\subsection{Expression of the rate function}

We consider a system of Ising spins on a two-dimensional square lattice
of linear size $L$, with periodic boundary conditions.
Thus, the number of spins $N$ is equal to $L^2$, and the number of bonds is equal to $2L^2$.
The energy (or Hamiltonian), for a given configuration of spins, reads
\beq\label{eq:energy}
E=-J\sum_{i=1,L}\sum_{j=1,L}(\s_{i,j}\s_{i,j+1}+\s_{i,j}\s_{i+1,j}).
\eeq
From now on we will set $J=1$, $k_B=1$ and denote the reduced coupling constant by $K=1/T$.

At each instant of time, a spin, denoted by $\s$, is chosen at random and flipped with a rate denoted for short by 
$w(\s)$.
We choose the following form of the rate function~\cite{gb2009,cg2013,gp2014}, where $\{\s_E,\s_N,\s_W,\s_S\}$ are the neighbours of the central spin $\s$, and E, N, W, S stand for East, North, etc.\footnote{The notation N for North should not be confused with the notation for the number of spins.}
\beqa
w(\s)&=&
\frac{\alpha}{2}\big[
1-\gamma\frac{1+V}{2}\s(\s_E+\s_N)-\gamma\frac{1-V}{2}\s(\s_W+\s_S)
\nonumber\\
&+&\gamma^2\frac{1+V}{2}\s_E\s_N+\gamma^2\frac{1-V}{2}\s_W\s_S
\big],
\label{eq:rate}
\eeqa
where $\alpha$ fixes the scale of time, $\gamma=\tanh 2K$ and $V,$ named the velocity, is the asymmetry (or irreversibility) parameter, which allows to interpolate between the symmetric case ($V=0$) and the totally asymmetric ones ($V=\pm1$).
This expression of the rate function satisfies the condition of global balance, that is to say, leads to a Gibbsian stationary measure with respect to the Hamiltonian~(\ref{eq:energy}) even if the condition of detailed balance is not satisfied\footnote{We refer the reader to~\cite{gb2009,cg2013} for a detailed account of this property.}.
It represents one, among many, possible expression of a rate function, for the kinetic Ising model on the square lattice, possessing this property.
This expression is invariant under up-down spin symmetry.
In~(\ref{eq:rate}), for $V>0$ (resp. $V<0$) the couple (N, E) (resp. (S, W)) is more influential on the central spin than the other one. 

The two particular cases of symmetric dynamics ($V=0$), and completely asymmetric dynamics ($V=\pm1$), deserve special attention.
First, if $V=0$, then the rate function
\beqa
w(\s)=\frac{\alpha}{2}\big[
1-\frac{\gamma}{2}\s(\s_E+\s_N+\s_W+\s_S)
+\frac{\gamma^2}{2}(\s_E\s_N+\s_W\s_S)
\big]
\nonumber\\
\label{eq:rate0}
\eeqa
satisfies the condition of detailed balance.
This form is different from the usual Glauber rate function
\beq\label{glau2D}
w(\s)=\frac{\alpha}{2}\big[1-\s\tanh [K(\s_E+\s_N+\s_W+\s_S)]\big],
\eeq
which can also be written in terms of spin operators as
\beqa\label{glau2D+}
w(\s)=\frac{\alpha}{2}\Big[1-\frac{\gamma(2+\gamma^2)}{4(1+\gamma^2)}\s(\s_E+\s_N+\s_W+\s_S)
\nonumber\\
+\frac{\gamma^3}{4(1+\gamma^2)}\s(\s_E\s_N\s_W+\s_N\s_W\s_S+\s_W\s_S\s_E+\s_S\s_E\s_N)\Big].
\eeqa
The Glauber rate function is fully symmetric under a permutation of the neighbouring spins, or, equivalently, only depends on the variation of the energy due to a flip, 
\beq
\Delta E=2\sigma (\sigma_E+\sigma_N+\sigma_W+\sigma_S).
\eeq
In contrast, the rate function~(\ref{eq:rate0}) with $V=0$ does not possess this property: it is not fully symmetric under a permutation of the neighbouring spins and does not depend on $\Delta E$ only.
In other words the neighbouring spins of the central spin do not all play equivalent roles, i.e., cannot be interchanged.
This can be demonstrated by looking at the values taken by~(\ref{eq:rate0}) according to the values of the local field $h=\s_E+\s_N+\s_W+\s_S$: if this sum is equal to $\pm2$ then all rates are the same.
In contrast, if the sum vanishes then the rate function takes two different values, $\alpha(1\pm\gamma^2)$, according to the configuration of the neighbours $\{\s_E,\s_N,\s_W,\s_S\}$: $\alpha(1+\gamma^2)$ corresponds to $\{++--\}$ or $\{--++\}$, while $\alpha(1-\gamma^2)$ corresponds to all other configurations.
This is a manifestation of an anisotropy in the dynamics.
By comparison, for the Glauber case, 
if the local field vanishes, the rate function takes only one value, $\alpha/2$.

Let us add a word of caveat on the terminology used in the present work.
As said above, the dynamics defined by the rate function~(\ref{eq:rate0}) is anisotropic.
Yet, for short, when speaking of asymmetric rates we will have in mind the case $V\ne0$.

\begin{table}[ht]
\caption{Properties of the dynamics investigated in the present work.
Glauber refers to~(\ref{glau2D}), $V=0$ to~(\ref{eq:rate0}), and $V\ne0$ to~(\ref{eq:rate}).}
\label{tab:1}
\begin{center}
\begin{tabular}{|c||c|c|c|}
\hline
&Glauber&$V=0$&$V\ne0$\\
\hline
symmetry&full&partial&no\\
isotropy&yes&no&no\\
directedness&no&no&yes\\
reversibility&yes&yes&no\\
\hline
\end{tabular}
\end{center}
\end{table}

The case with $V=1$ corresponds to the totally asymmetric dynamics where the central spin $\s$ is influenced by its East and North neighbours only.
Then
\beq\label{eq:ne}
w(\s)=\frac{\alpha}{2}
\left[1-\gamma\s(\s_E+\s_N)+\gamma^2\,\s_E\s_N\right].
\eeq
A similar expression, involving $\s_W$ and $\s_S$, holds for $V=-1$:
\beq\label{eq:sw}
w(\s)=\frac{\alpha}{2}
\left[1-\gamma\s(\s_W+\s_S)+\gamma^2\,\s_W\s_S\right].
\eeq
A remarkable fact about these expressions is that they are {\it unique}, up to the scale of time fixed by the choice of $\alpha$, in the following sense.
For rate functions only involving NEC (North, East, Central) spins, or SWC (South, West, Central) spins,
i.e., when only two neighbours, chosen among the four possible ones, are influential upon the central spin,
imposing the condition of global balance uniquely determines the expressions~(\ref{eq:ne}) or (\ref{eq:sw})~\cite{gb2009,cg2013}.
The rate function~(\ref{eq:ne}) for the totally asymmetric NE dynamics was originally given in~\cite{kun}, under a slightly different form, but without considerations on its derivation or its unicity.

Table~\ref{tab:1} summarizes the properties of the dynamics investigated in the present work.

As a final remark, let us point out that the rate function~(\ref{eq:rate}) is a linear combination of~(\ref{eq:ne}) and of~(\ref{eq:sw}), which makes its definition rather natural.
Of course, one could define a more symmetrical rate function by taking a linear superposition of the expressions corresponding to the four couples NE~(\ref{eq:ne}), NW, SW~(\ref{eq:sw}), and SE.
This choice would however imply the presence of more than one irreversibility parameter in the rate function.

\subsection{Rules of zero-temperature dynamics}
\label{sec:rules}

\begin{figure}[ht]
\begin{center}
\includegraphics[angle=0,width=0.5\linewidth]{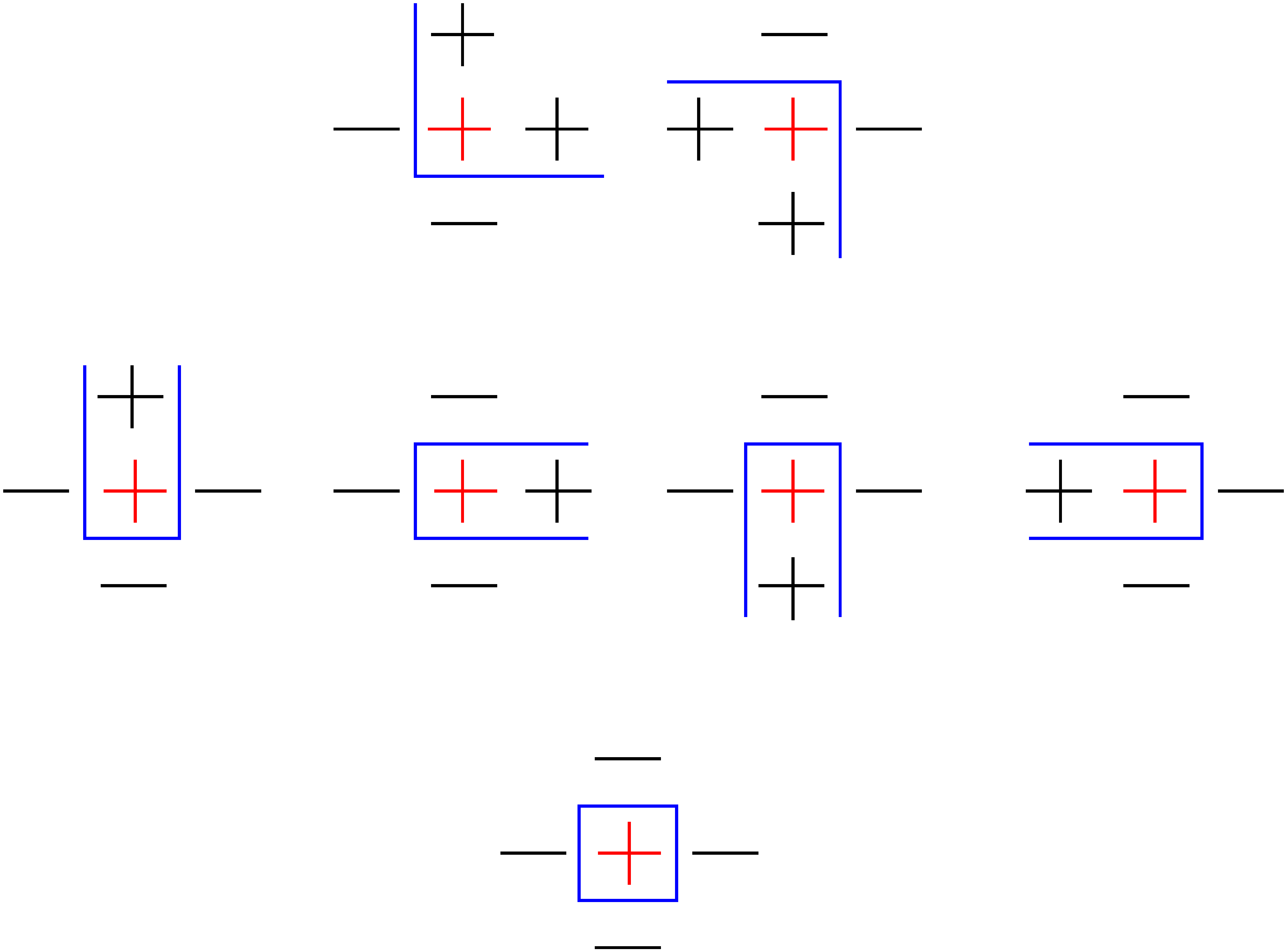}
\end{center}
\caption{Local configurations with non-vanishing flipping rates for the central spin (in red)
in the directed Ising model with rate function~(\ref{eq:rate}) at $T=0$.
The blue segments represent the domain walls.
The first line corresponds to a value of the local field $h=0$, the second one to $h=-2$, the third one to $h=-4$.
The associated rates are, from left to right, in units of $\alpha$,
first line: $1-V$, $1+V$ (Glauber: 1/2, 1/2);
second line: $1-V$, $1-V$, $1+V$, $1+V$ (Glauber: 1, 1, 1, 1);
third line: $2$ (Glauber: 1).
}
\label{fig:moves}
\end{figure}

\begin{figure}[ht]
\begin{center}
\includegraphics[angle=0,width=0.4\linewidth]{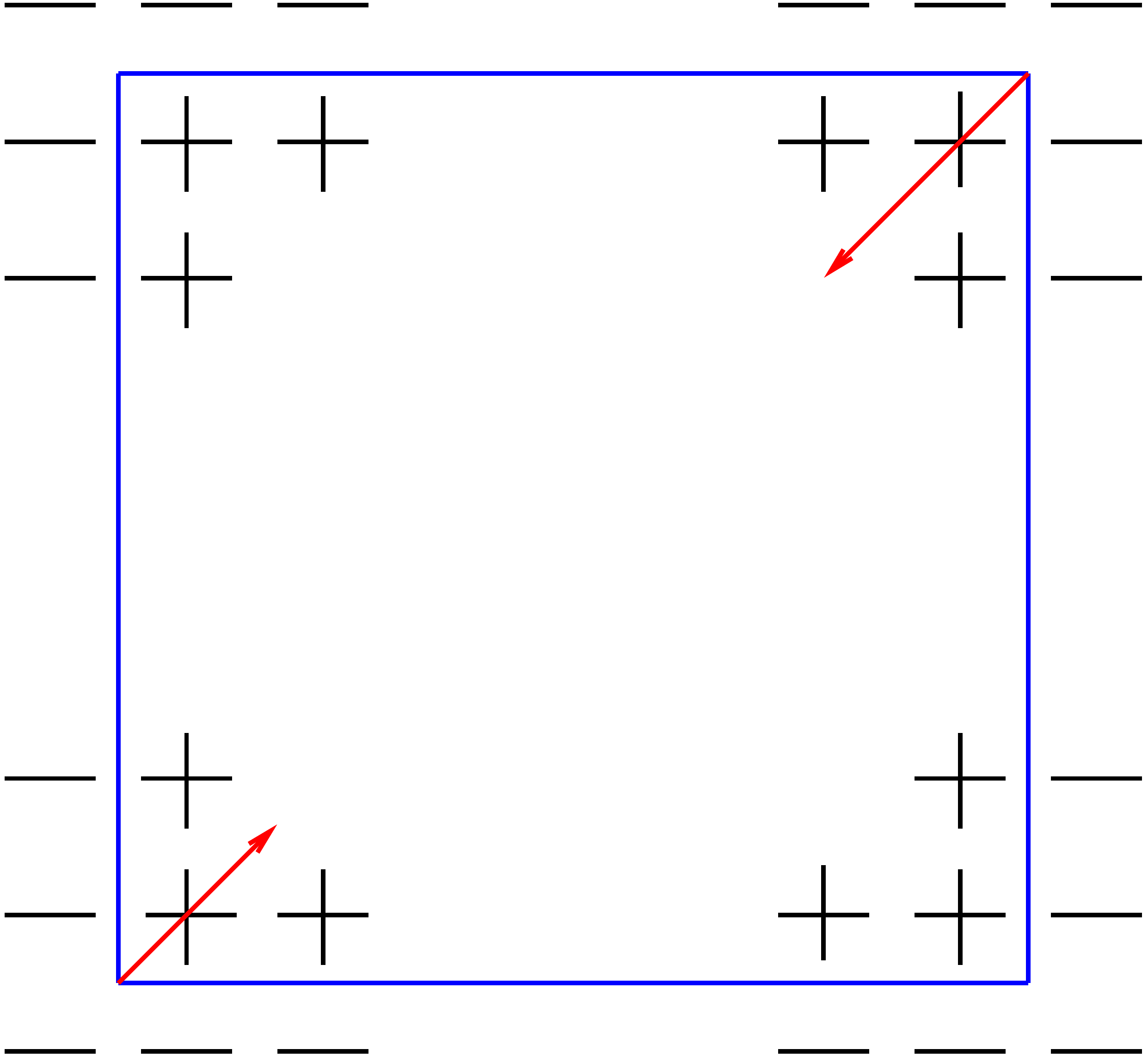}
\end{center}
\caption{Possible moves for a minority ($+$) square for the directed Ising model with rate function~(\ref{eq:rate}) at $T=0$.
Only the North-East and South-West corners move.
The big arrow corresponds to a rate equal to $1+V$ (in units of $\alpha$), the smaller one to $1-V$ (see the first line of figure~\ref{fig:moves}).
The same holds for a minority ($-$) square.
For Glauber dynamics the flipping rates are the same for the four corners.
}
\label{fig:moves2}
\end{figure}

\begin{figure}[ht]
\begin{center}
\includegraphics[angle=0,width=0.4\linewidth]{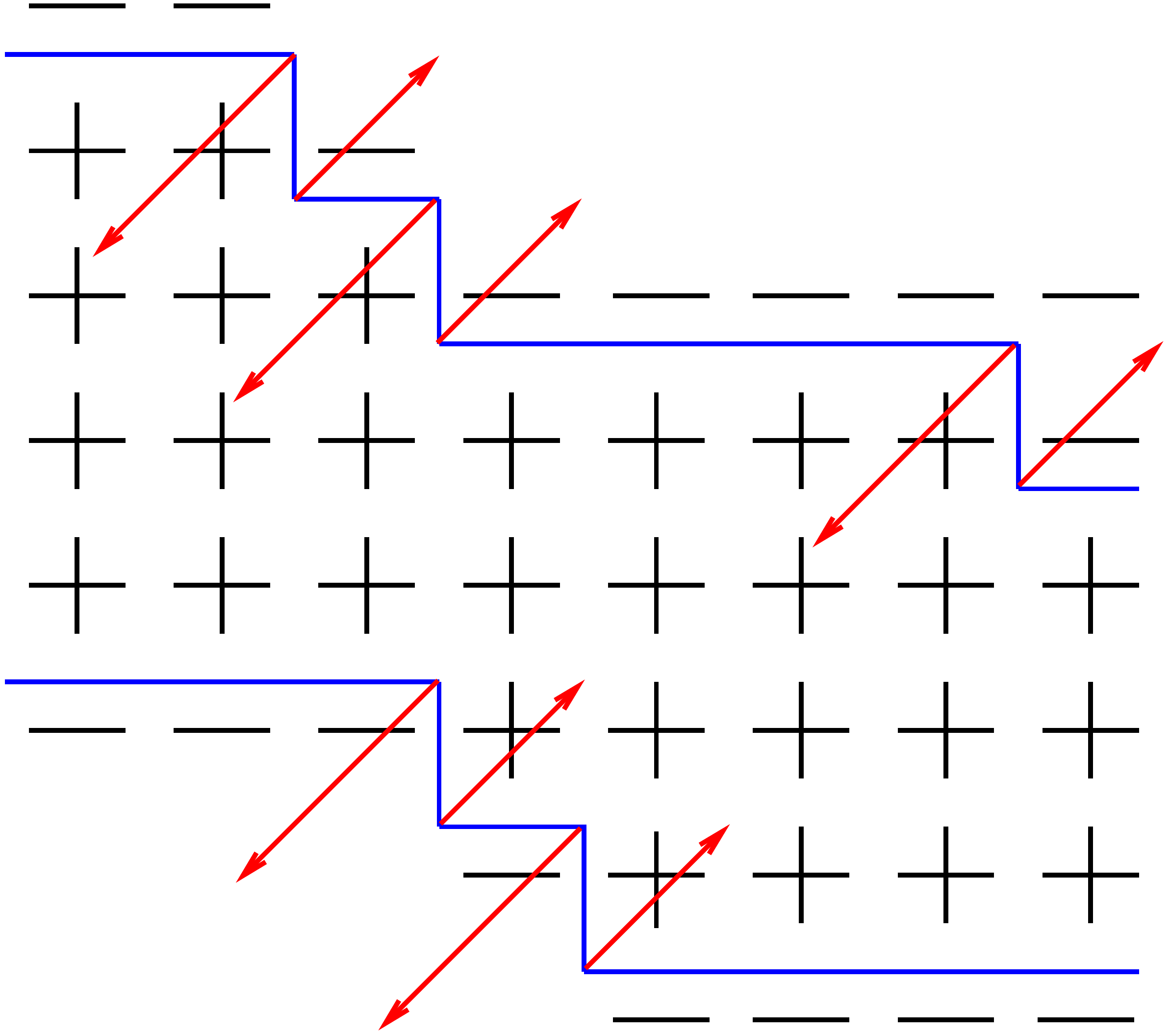}
\end{center}
\caption{Evolution of a stripe oriented perpendicular to the North-East direction for the directed Ising model with rate function~(\ref{eq:rate}) at $T=0$.
Stripes oriented parallel to the North-East direction are blocked while they evolve for Glauber dynamics. 
}
\label{fig:moves3}
\end{figure}

Let us first remind that for zero-temperature Glauber dynamics
moves such that $\Delta E>0$ have zero rates,
moves such that $\Delta E=0$ have rate $1/2$ and those leading to $\Delta E<0$ have rate $1$, in units of $\alpha$  (see~(\ref{glau2D}) with $\gamma=1$).

In the present case the non-vanishing rates at zero temperature  are read off from~(\ref{eq:rate}) and correspond to configurations pictured in figure~\ref{fig:moves}.
The first line correspond to a local field $h=0$ ($\Delta E=0$), the second line to $h=-2$ ($\Delta E<0$), the third one to $h=-4$ ($\Delta E<0$).
In units of $\alpha$, and reading from left to right, the rates are equal to $1-V$ and $1+V$ if $h=0$, to $1-V$, $1-V$, $1+V$, $1+V$, if $h=-2$, and to $2$, if $h=-4$.
Due to spin-reversal symmetry, the rates are unchanged if one reverts the signs of all spins in figure~\ref{fig:moves}.

There is a notable difference between Glauber dynamics and the dynamics with $V=0$, namely that 
for the former all configurations with $h=0$ have rates equal to $1/2$,
while, for the latter, configurations with $h=0$ where the restricted local field $\s_E+\s_N=0$ (and therefore $\s_W+\s_S=0$) have zero rates.

The zero-temperature dynamics with $V=1$ (resp. $V=-1$) is very simple 
since only one single non-zero rate remains, corresponding to the configurations where the North and East spins (resp. South and West spins) are both negative.
(For $V=-1$, the same holds for South and West spins.)

Figure~\ref{fig:moves2} depicts the possible moves of a minority $(+)$ square with the rate function~(\ref{eq:rate}).
Only the North-East and South-West corners can move, as can be seen from the first line of figure~\ref{fig:moves}.
We thus infer that a stripe oriented parallel to the North-East direction will not move, even when $V=0$, whereas a stripe oriented perpendicular to the North-East direction will, as depicted in figure~\ref{fig:moves3}.
Examples of such configurations will be encountered and commented below (see figures~\ref{fig:snapp} and~\ref{fig:snapp2}).

\subsection{How to compare two dynamics?}

Since in the course of this work we shall often compare the results obtained with the rate function~(\ref{eq:rate}) to those obtained with Glauber rate function~(\ref{glau2D}), we now discuss the choice of the time-scale parameter $\alpha$ made in our simulations.

At infinite temperature ($\gamma=0$) the two expressions~(\ref{eq:rate}) and~(\ref{glau2D}) yield rates equal to $\alpha/2$,
hence the two time scales of these dynamics are identical.
On the other hand, in a simulation at finite (or zero) temperature, if, for practical purposes, one wants the rates to be less than 1, one should fix the value of $\alpha$ according to the largest rate, {\it simultaneously for both rates}.
The largest rates are
\beqa
{\rm Glauber}\,: \frac{\alpha}{2}(1 +\gamma)^2/(1 +\gamma^2) &&\zeq{T\to0}  \frac{\alpha}{2}\times 2,
\nonumber\\
{\rm Directed}:\frac{\alpha}{2}(1+\gamma)^2 &&\zeq{T\to0}  \frac{\alpha}{2}\times 4.
\eeqa
In the simulations presented hereafter the choice $\alpha=1/2$ has been done for both rates.

\section{Lifetime of a minority cluster}
\label{sec:lifetime}

\begin{figure}[ht]
\begin{center}
\includegraphics[angle=0,width=0.9\linewidth]{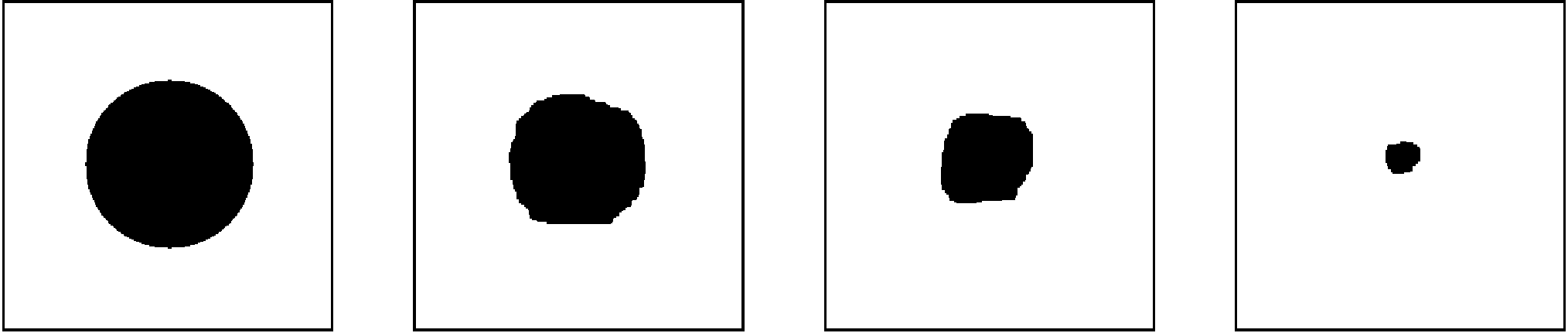}
\end{center}
\caption{Snapshots showing the disappearance of a circle of minority spins when using Glauber dynamics~(\ref{glau2D}) at $T=0$. 
The system size is 
$L=300$, and the radius of the circle $a=L/4$.
Times are $t=0$, $t^{*}/3$, $2t^{*}/3
$, $0.95 t^{*}$, where $t^{*}=A=\pi(300/4)^2\approx 17671$.}
\label{fig:circleGlauber}
\end{figure}

\begin{figure}[ht]
\begin{center}
\includegraphics[angle=0,width=0.9\linewidth]{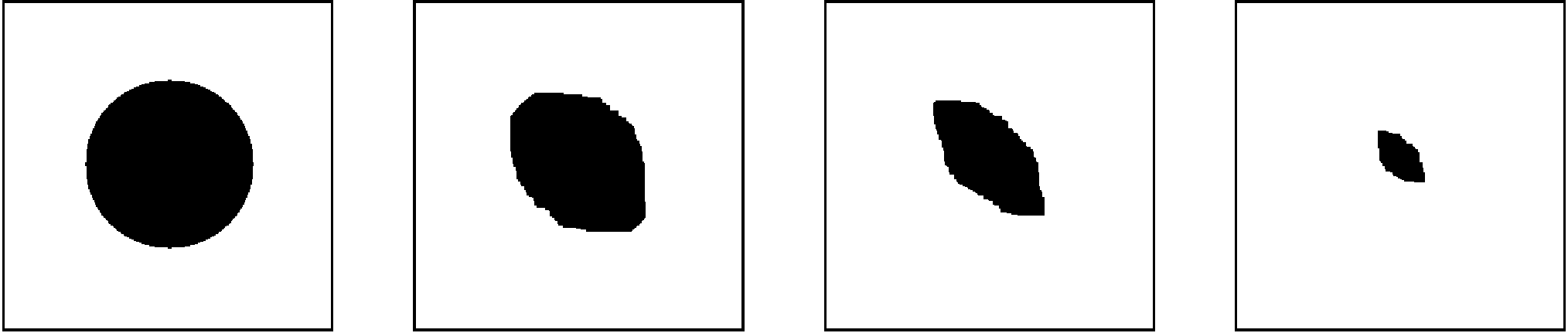}
\end{center}
\caption{
Snapshots showing the disappearance of a circle of minority spins for reversible rates~(\ref{eq:rate0}) with $V=0$ at $T=0$.
System size and times are as in figure~\ref{fig:circleGlauber}.}
\label{fig:circleV0}
\end{figure}

\begin{figure}[ht]
\begin{center}
\includegraphics[angle=0,width=0.9\linewidth]{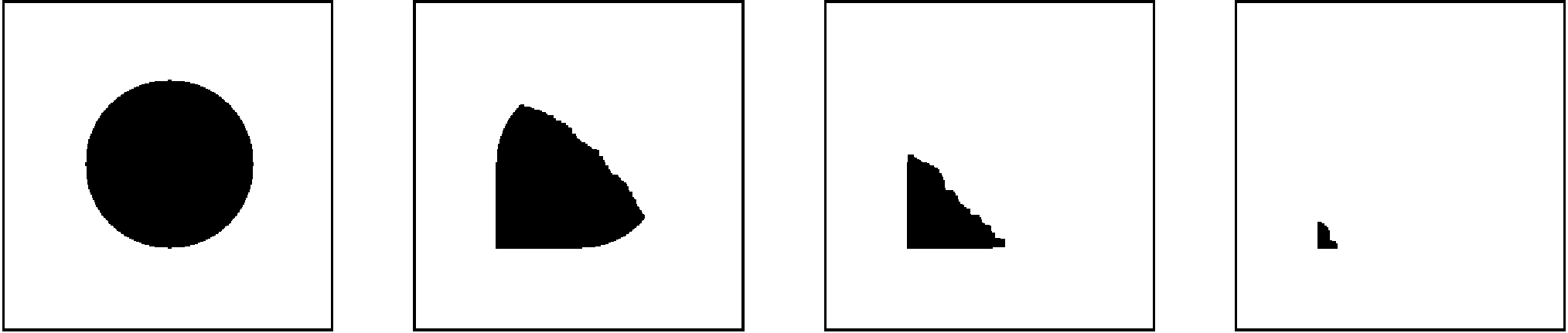}
\end{center}
\caption{Snapshots showing the disappearance of a circle of minority spins for totally asymmetric rates~(\ref{eq:ne}) with $V=1$ at $T=0$.
The circle deforms quickly into a right triangle. 
Here $t^{*}=500$ which is the approximate value of $\tcluster$ extrapolated from figure~\ref{fig:lifetime2} for $\sqrt{A}=\sqrt{\pi}(300/4)\approx133$.
}
\label{fig:circleV1}
\end{figure}

\begin{figure}[ht]
\begin{center}
\includegraphics[angle=0,width=0.9\linewidth]{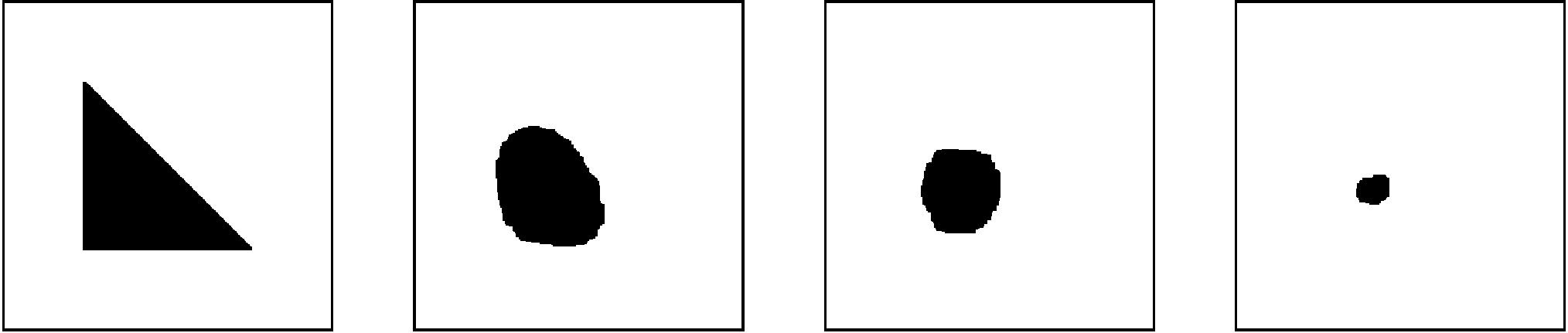}
\end{center}
\caption{Snapshots showing the disappearance of a right triangle of minority spins when using Glauber dynamics~(\ref{glau2D}) at $T=0$. 
The system size is 
$L=300$, and the side of the triangle $a=L/2$.
Here $t^{*}=A=L^2/8=11250$. 
}
\label{fig:triangleGlauber}
\end{figure}

\begin{figure}[ht]
\begin{center}
\includegraphics[angle=0,width=0.9\linewidth]{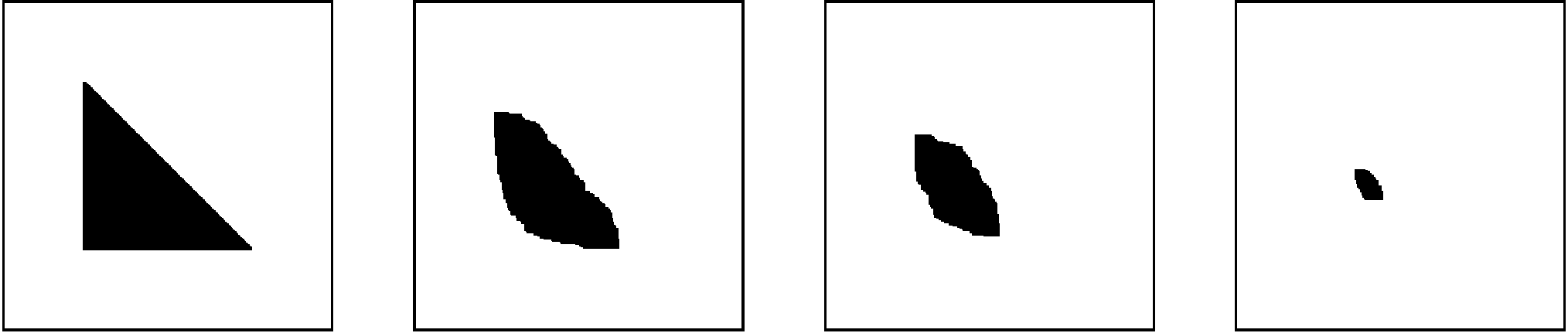}
\end{center}
\caption{
Snapshots showing the disappearance of a right triangle of minority spins for reversible rates~(\ref{eq:rate0}) with $V=0$ at $T=0$.
System size and times are as in figure~\ref{fig:triangleGlauber}.
}
\label{fig:triangleV0}
\end{figure}

\begin{figure}[ht]
\begin{center}
\includegraphics[angle=0,width=0.9\linewidth]{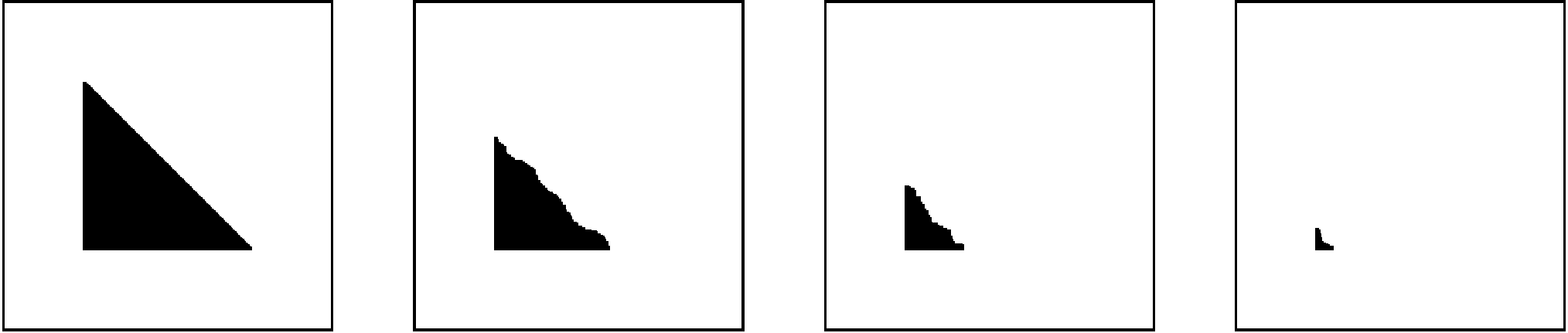}
\end{center}
\caption{Snapshots showing the disappearance of a right triangle of minority spins for totally asymmetric rates~(\ref{eq:ne}) with $V=1$ at $T=0$.
Here $t^{*}=290$ 
which is the approximate value of $\tcluster$ extrapolated from figure~\ref{fig:lifetime2} for $\sqrt{A}=(300/4\sqrt{2})\approx106$.
}
\label{fig:triangleV1}
\end{figure}

A first approach to the determination of the scales of time occurring in the coarsening process is provided by analyzing the lifetime of a ($+$) cluster in a ($-$) sea, as depicted in~figures~\ref{fig:circleGlauber}-\ref{fig:triangleV1}.

Figures~\ref{fig:circleGlauber}-\ref{fig:circleV1} show the fate of a circle under Glauber dynamics, $V=0$ dynamics, and $V=1$ dynamics, respectively.
The two first figures show that the scale of time for the disappearance of the circle is the same for Glauber and $V=0$ dynamics, but that an anisotropy is present in the latter case, while in the former the circle shrinks isotropically.
Figure~\ref{fig:circleV1} shows that, when $V=1$, the scale of time for the disappearance of the circle is much shorter than for the two reversible dynamics of figures~\ref{fig:circleGlauber} and~\ref{fig:circleV0}, 
and that the circle is driven to the shape of a right triangle.
Figures~\ref{fig:triangleGlauber}-\ref{fig:triangleV1} show that the final stages of the evolution of a right triangle are the same as for a circle.

\begin{figure}[ht]
\begin{center}
\includegraphics[angle=0,width=0.6\linewidth]{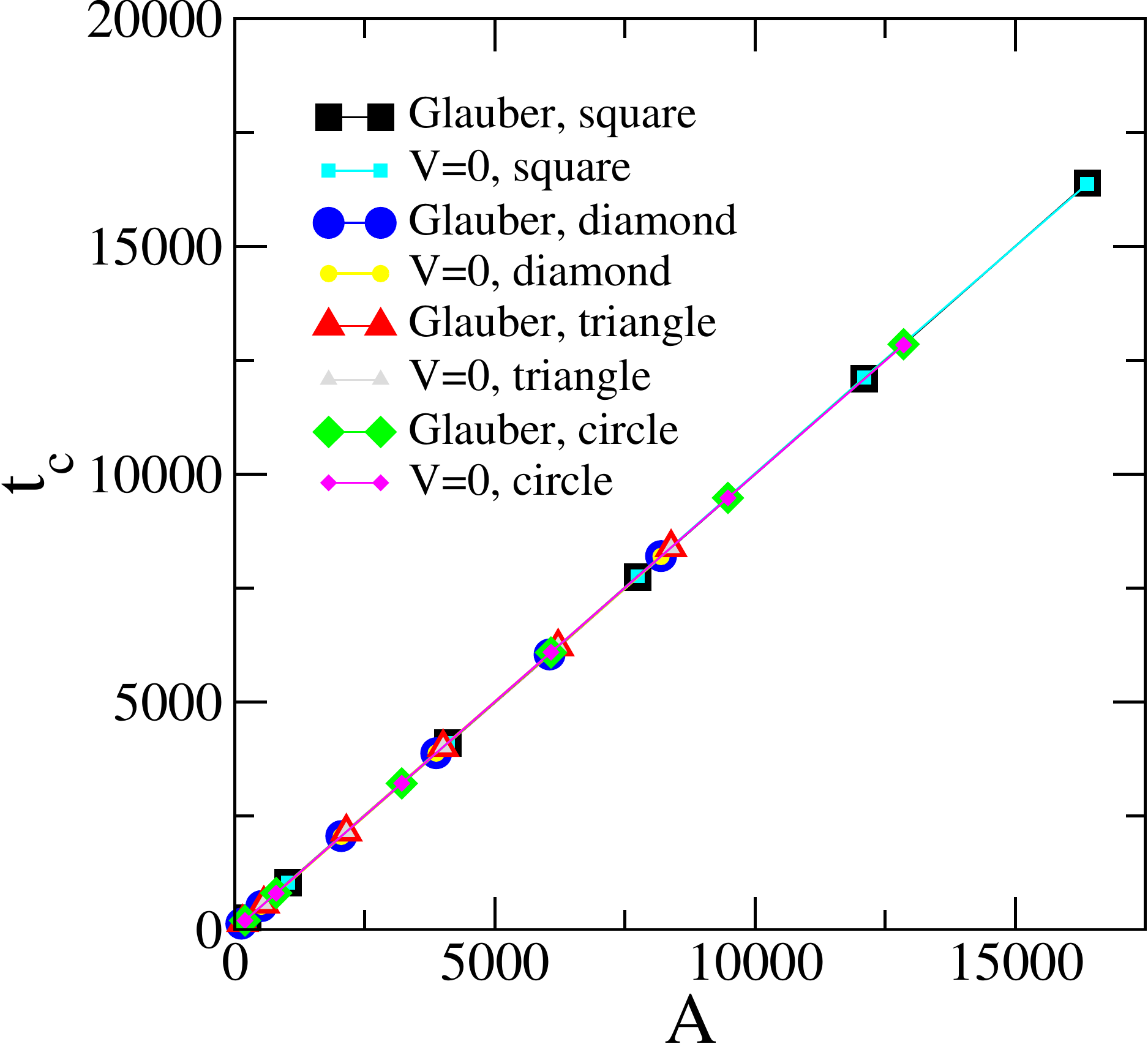}
\end{center}
\caption{Mean lifetime $\tcluster$ of a cluster of minority spins in a sea of majority spins as a function of area for Glauber
dynamics as well as for the reversible dynamics~(\ref{eq:rate0}) with $V=0$ at $T=0$. 
The data result from averaging over 8000 independent histories. 
}
\label{fig:lifetime1}
\end{figure}

\begin{figure}[ht]
\begin{center}
\includegraphics[angle=0,width=0.6\linewidth]{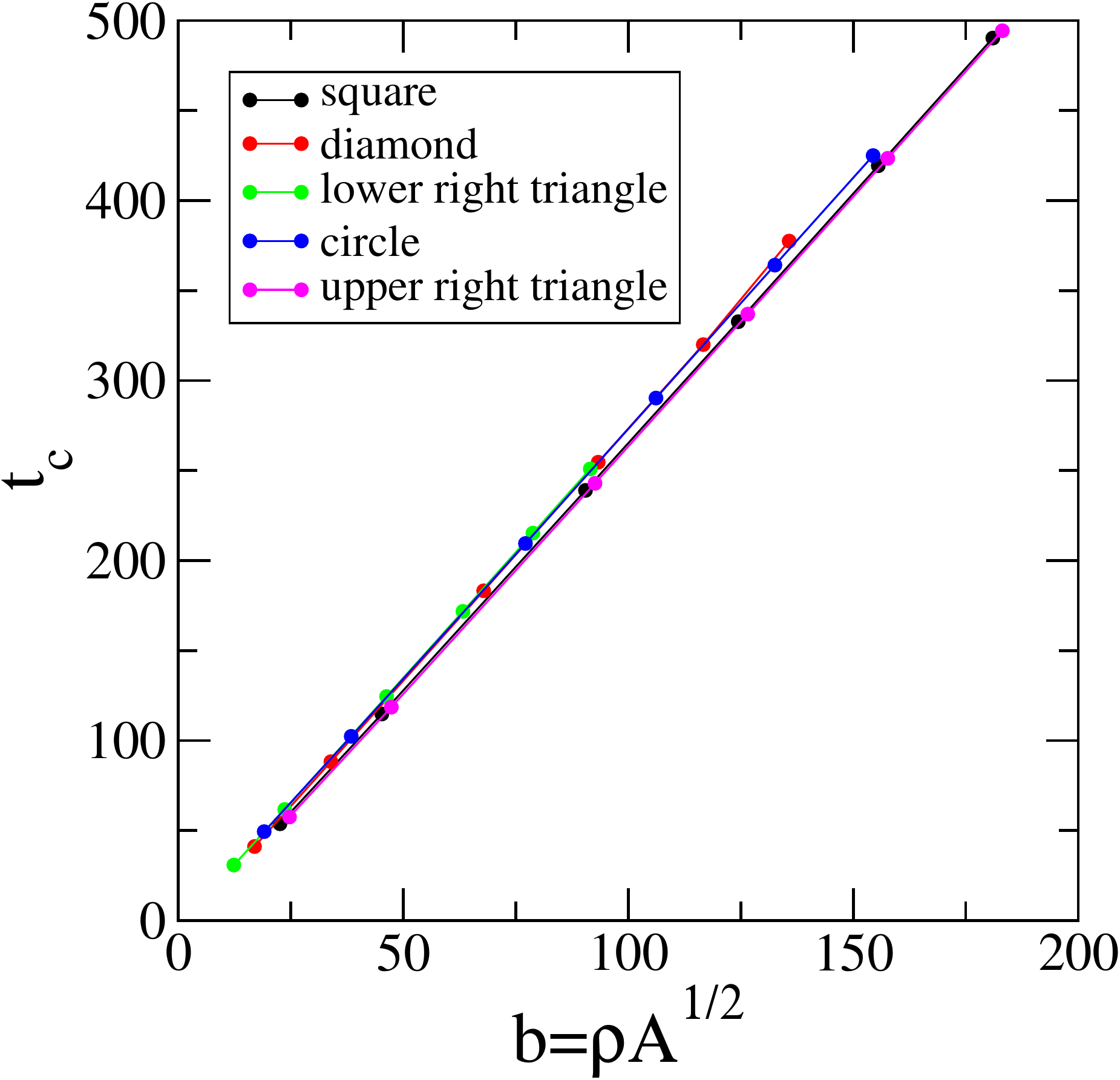}
\end{center}
\caption{Mean lifetime of a cluster of minority spins in a sea of majority spins as a function of the length $b$ 
(see text and figure~\ref{fig:defb}) for
asymmetric dynamics with $V=1$ at $T=0$.
The data result from averaging over 8000 independent histories.
}
\label{fig:lifetime2}
\end{figure}

Figures~\ref{fig:lifetime1} and~\ref{fig:lifetime2} give the average time $\tcluster$ needed for a cluster to disappear.
As demonstrated by these figures,
there are two scales of time according to the value of $V$.
For reversible dynamics, $\tcluster$ scales as the area $A$ of the cluster, while as soon as $V$ is positive this time is proportional to $\sqrt{A}$.
This gives a first hint that asymmetric dynamics is ballistic, a feature which will be confirmed in the sequel by several methods.
We observe on figure~\ref{fig:lifetime1} that the asymptotic slope of $\tcluster$ against $A$ is the same for all clusters, with common value unity (with the choice made here of $\alpha=1/2$).

\begin{figure}[ht]
\begin{center}
\includegraphics[angle=0,width=0.7\linewidth]{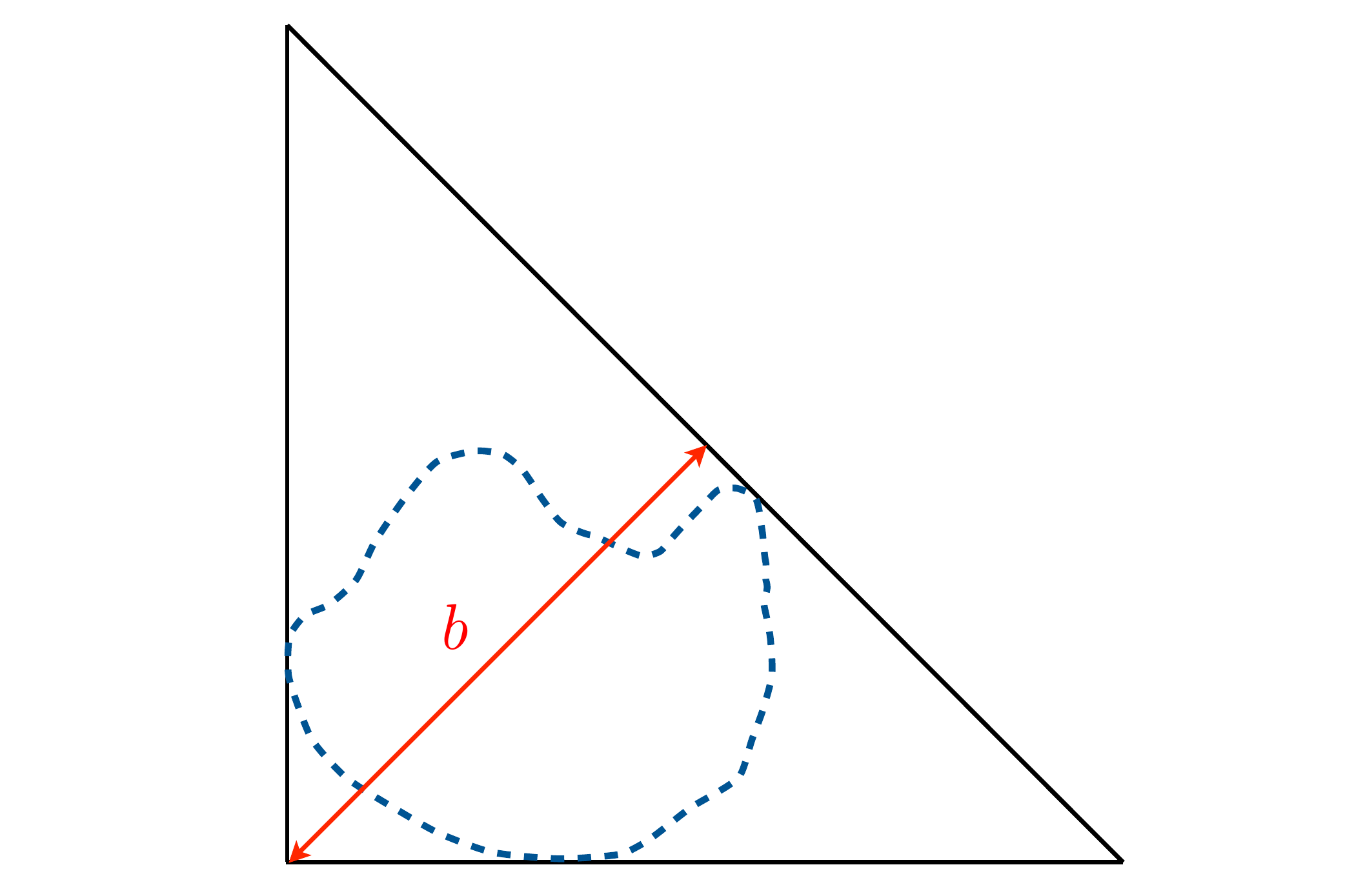}
\end{center}
\caption{Cluster inscribed in a right triangle and definition of length $b$.
}
\label{fig:defb}
\end{figure}

In contrast,  if $V=1$, the slope of $\tcluster$ against $\sqrt{A}$ depends on the shape of the cluster.
However a simple rescaling yields figure~\ref{fig:lifetime2}, as follows.
Consider a $(+)$ cluster in a sea of $(-)$ spins.
It can be inscribed in a right triangle, as depicted in figure~\ref{fig:defb}.
The tip of the South-West right angle of the triangle turns out to be
the South-West corner point appearing in the snapshots.
Let us denote by $b$ the distance between this point and the hypotenuse of the right triangle.
Clusters sharing the same value of $b$ should shrink equally rapidly.
Hence, in order to compare clusters of different shapes we plot $t_c$ against $b=\rho\sqrt{{A}}$, where $A$ is the area, and $\rho$ the scale factor.
For instance for a right triangle (lower half of a square with sides of length $a$ parallel to the axes), $A=a^2/2$, $b=a/\sqrt{2}$, thus $\rho=1$.
Likewise for the above mentioned square, $A=a^2$, $b=a\sqrt{2}$, hence $\rho=\sqrt{2}$.
For a circle of radius $a$, $\rho=(1+\sqrt{2})/\sqrt{\pi}$, for the diamond, $\rho=3/2$, and for the upper right triangle, $\rho=2$.
We thus observe on figure~\ref{fig:lifetime2} that the asymptotic slope of $\tcluster$ against $b$ is the same for all clusters, with common value very close to $2\sqrt{2}$.
More generally we observe that
\beq\label{eq:tc}
V t_c\approx\frac{\sqrt{2}\,b}{\alpha},
\eeq
for any $V\ne0$ and for $\alpha$ arbitrary.

The time evolution of a shrinking domain under Glauber dynamics has been addressed in~\cite{domany,chayes} and we refer the reader to these references for an explanation of the slope of $t_c$ against $b$ observed in figure~\ref{fig:lifetime1}.\footnote{More recent developments on related subjects can be found in~\cite{krap1,krap2,krap3,krap4,karma}.}  
For $V=0$ dynamics, since the slope is unchanged, it is clear that the dynamics of the cluster is dominated by diffusive moves, corresponding to the first line in figure~\ref{fig:moves}. 

We now present a simple argument in order to explain the above observations on figure~\ref{fig:lifetime2} for the case where $V\ne0$.
We define a speed $v$ (to be distinguished from the so-called velocity $V$) by $b=v t_c$.
Let us consider an infinite interface oriented in the North-West to South-East direction made of a regular succession of horizontal and vertical domain walls.
As is well known, one can describe the evolution of this interface by the asymmetric simple exclusion process (ASEP) on a line as follows.
An horizontal domain wall corresponds to a particle, a vertical domain wall to a hole.
The current of particles is $J=(p-q)\rho(1-\rho)$ where $p$ and $q$ are the rates for a jump of the particle, respectively to the right or to the left and $\rho$ is the density of particles.
Here $p=\alpha(1+V)$ and $q=\alpha(1-V)$, hence $J=\alpha V/2$ since $\rho=1/2$ in the present case.
In a lapse of time $\tau$ such that each particle has advanced by one step, corresponding to a displacement of the interface in the South-West direction equal to $\sqrt{2}/2$,
the total current of particles through a bond in the ASEP is equal to $J\tau=\rho$.
Hence $v=\sqrt{2}/(2\tau)$ and finally
\beq
v=\frac{\alpha V}{\sqrt{2}}.
\eeq
This simple argument, which accounts for~(\ref{eq:tc}), demonstrates that the bulk of the phenomenon comes from the ballistic motion of the interface oriented in the North-West to South-East direction.

\begin{figure}[ht]
\begin{center}
\includegraphics[angle=0,width=0.5\linewidth]{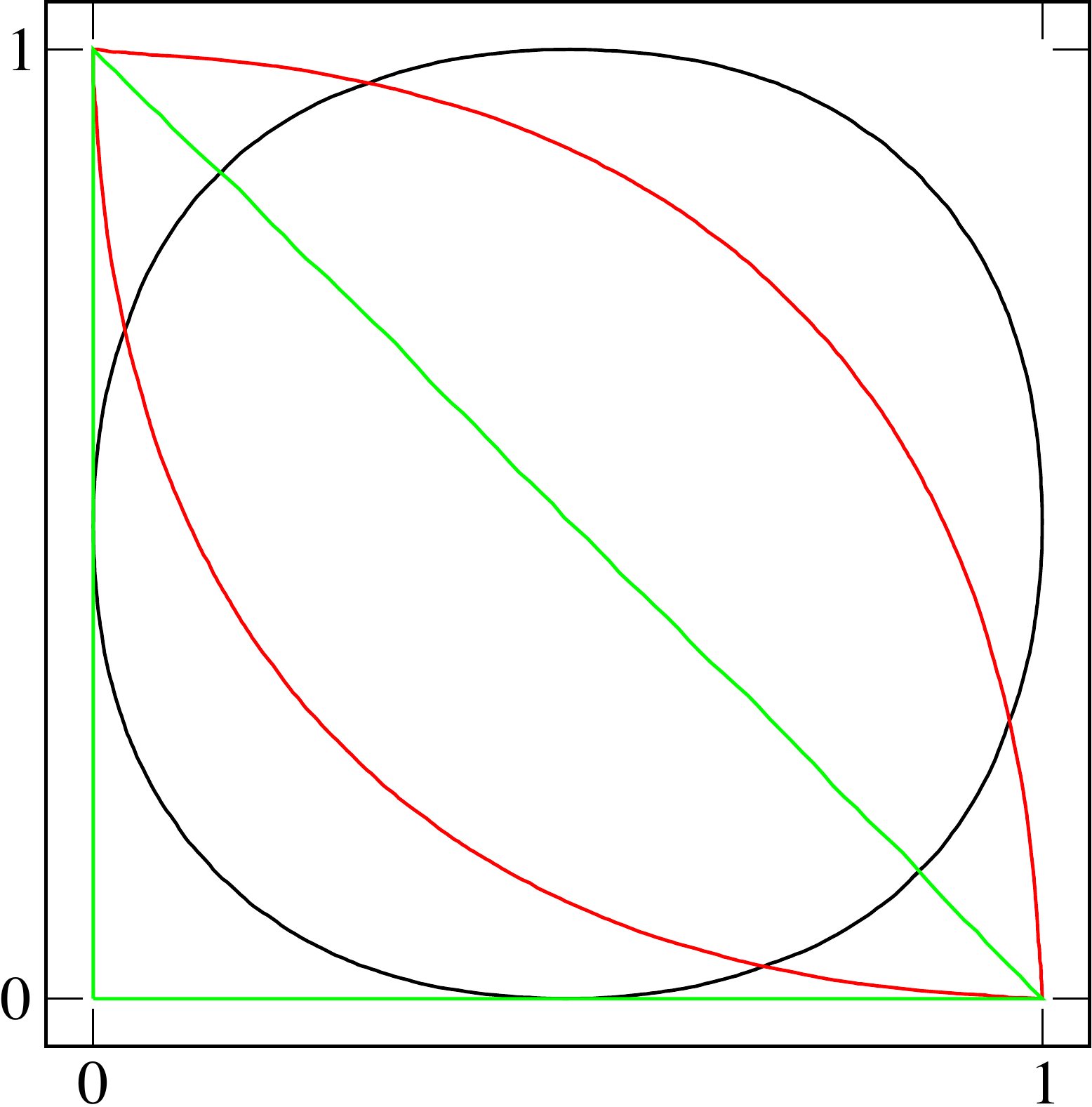}
\end{center}
\caption{Limiting shapes of a cluster for three dynamics. Glauber: black circular-like curve, $V=0$: red lens, $V=1$: green triangle. 
}
\label{fig:limiting}
\end{figure}

\begin{figure}[ht]
\begin{center}
\includegraphics[angle=0,width=0.5\linewidth]{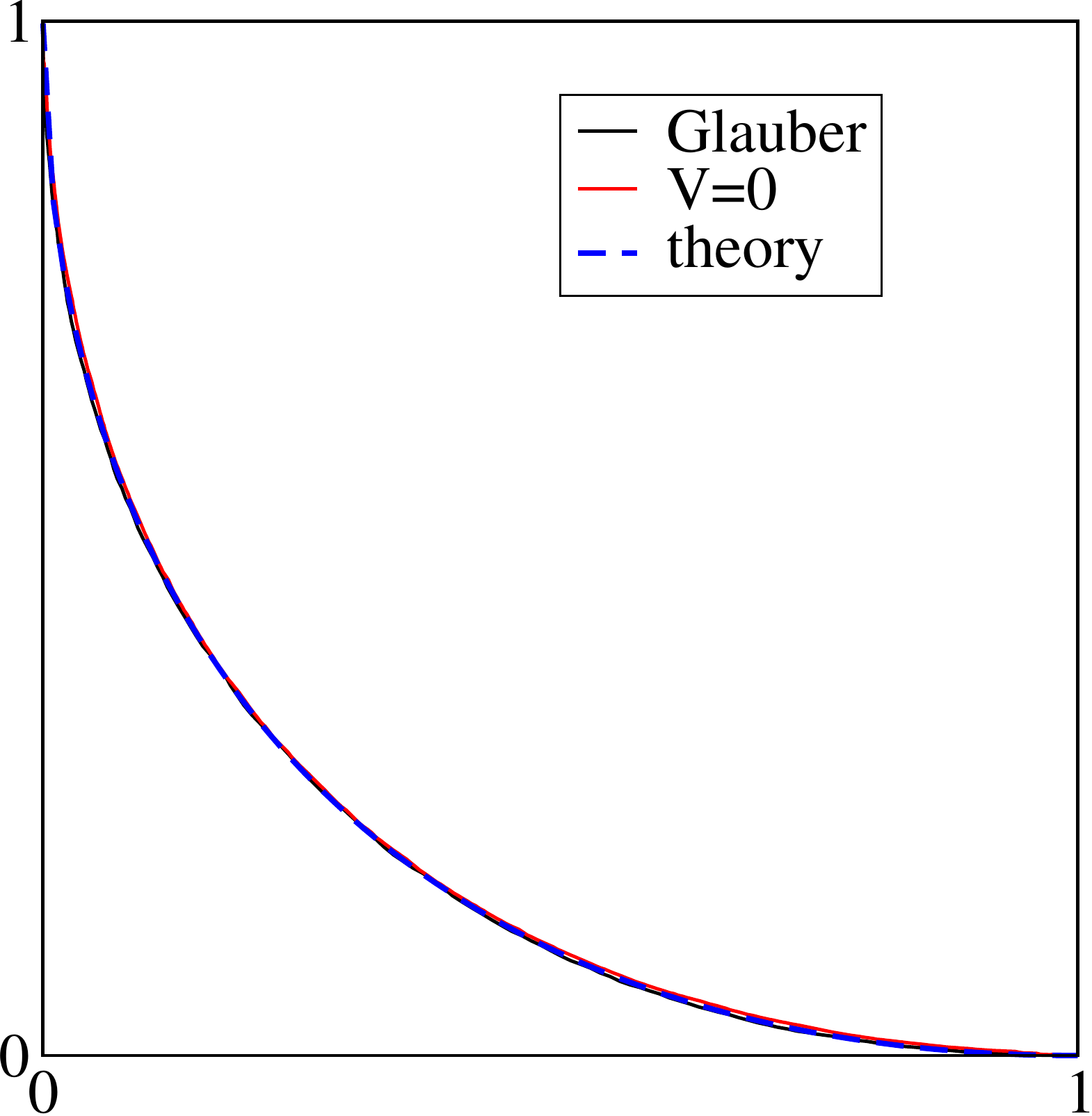}
\end{center}
\caption{Parts of the limiting shapes of figure~\ref{fig:limiting} for Glauber and $V=0$ dynamics after rescaling the former by a linear factor equal to 2.
The dashed blue curve is the theoretical prediction obtained in~\cite{krap2,karma}.
}
\label{fig:limiting2}
\end{figure}

To close, let us mention that rescaling the size of the cluster by the relevant time scale ($\sqrt{t}$ for reversible dynamics, or $t$ for irreversible dynamics) yields limiting shapes for the shrinking clusters.
This has been thoroughly investigated for Glauber dynamics~\cite{krap2,karma}.
We report our results for the $V=0$ and $V=1$ dynamics in figure~\ref{fig:limiting} together with those for Glauber dynamics.
The limiting shape of a cluster for the $V=1$ dynamics is a triangle for obvious reasons.
More interestingly, the limiting shape for the $V=0$ dynamics is the same as that for Glauber dynamics (after rescaling the latter by a linear factor equal to 2).
In figure~\ref{fig:limiting2} the theoretical prediction~\cite{krap2,karma} is represented by a dashed curve, indistinguishable from the two other curves obtained by simulations.
The identity between the limiting curves of Glauber and $V=0$ dynamics stems from the fact that, in the long-time limit, the only relevant moves for the evolution of a cluster are those corresponding to the first line of figure~\ref{fig:moves}.

\section{Energy at $T=0$}
\label{sec:energy}

In this section we start the investigation of the scales of time for a finite system undergoing phase ordering after a quench from infinite temperature down to zero temperature.
We address the following questions: what can be said on the relaxation of the system to the ground state, or possibly to other configurations of higher energy?
What is the influence of the velocity?
This study will confirm the first set of observations made in the previous section.

We monitor the evolution of the system by measuring (minus) the energy of a bond, or density of energy, denoted by $\en(t)$.
When the system reaches the ground state the final value of $\en(t)$ is equal to one, therefore its convergence in the late-time regime to smaller values is the signature of blocked (or long-lived metastable) configurations.
We start by considering $\en(t)$ for the simple case of a one-dimensional chain of spins, as a preparation for the study of the two-dimensional system, that we consider next.

\subsection{Linear chain on a ring}

Let the system size be $L$, which is also the number of spins $N$ in the chain, as well as the number of bonds.
The energy of the chain, for a given configuration of spins, reads
\beq
E=-\sum_{i}\s_{i}\s_{i+1},
\eeq
thus (minus) the energy per bond reads
\beq
\en=-\frac{E}{L}.
\eeq
In the ground state (all spins $+$ or all spins $-$), we have $E=-L$ hence
$\en=1$.
More generally we have $E=-L+2n_{\un}$ where $n_{\un}$ is the number of unsatisfied bonds (or domain walls),
because
the cost in energy for the creation of a domain wall in the system is equal to 2 units,
hence
\beq\label{eq:e1Dstat}
\en=1-\frac{2n_{\un}}{L}.
\eeq
For instance the excited state of lowest energy for a linear chain on a ring corresponds to a pair of domain walls, for which we have $E=-L+4$, and
$\en=1-4/L$.

Equation~(\ref{eq:e1Dstat}) still holds during the temporal evolution of the system.
The inverse fraction of unsatisfied bonds $n_{+-}/L$, which is the
typical distance between the domain walls, gives a natural definition of
a growing length $\xi_{E}(t)$ as
\beq
\xi_{E}(t)=\frac{L}{n_{+-}}.
\eeq
Thus finally
\beq\label{eq:1Dlt}
\en(t)=1-\frac{2}{\xi_{E}(t)}.
\eeq

\subsection{Two-dimensional system of spins on the square lattice}

The expressions above generalize easily to the two-dimensional square lattice.
The energy is given in~(\ref{eq:energy}).
The energy per bond is now given (in absolute value) by
\beq
\en=-\frac{E}{2L^2}.
\eeq
In the ground state, $E=-2L^2$, hence $\en=1$.
More generally we have $E=-2L^2+2n_{\un}$ where $n_{\un}$ is the number of unsatisfied bonds,
hence
\beq\label{eq:e2D}
\en=1-\frac{n_{\un}}{L^2}.
\eeq
Consider for example a configuration of the system with one (vertical or horizontal) stripe.
The number of unsatisfied bonds is equal to $n_{\un}=2 L$,
hence
\beq
\en=1-\frac{2}{L}\quad ({\rm horizontal\ or\ vertical\ stripe}).
\eeq
More generally, more complex structures will be characterized by
\beq\label{eq:complex}
\en_k=1-\frac{2k}{L},
\eeq
where $k=1,2,\dots$.
Such structures, as in figures~\ref{fig:snapp} and~\ref{fig:snapp2}, will be encountered in the sequel.

\begin{figure}[ht]
\begin{center}
\includegraphics[angle=0,width=0.7\linewidth]{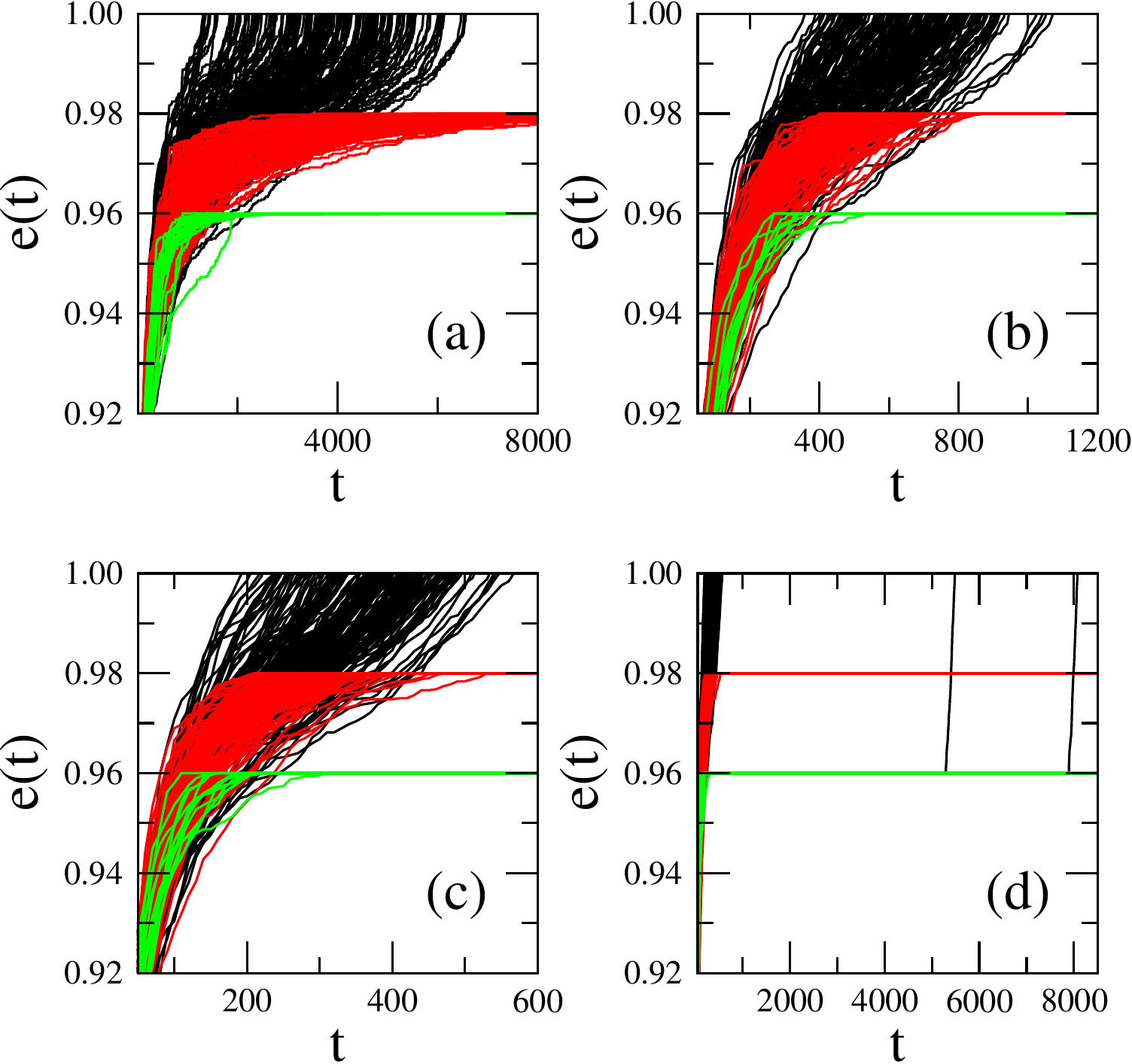}
\end{center}
\caption{300 histories for $\en(t)$ with (a): $V=0$, (b): $V=0.5$, (c): $V=1$ for a system of linear size $L=100$.
Figure (d) also depicts the $V=1$ case, but on another scale of time in order to exhibit trajectories jumping from $\en_2=0.96$ to $\en_0=1$.
See figure~\ref{fig:snapp} (right panel) for a snapshot of the system just before the jump of one of the two time-trajectories from 0.96 to 1 in figure (d).
}
\label{fig:e}
\end{figure}

During the temporal evolution of the system, (\ref{eq:e2D}) still holds, and, as for the chain, we define the growing length $\xi_{E}(t)$ as the inverse of the fraction of unsatisfied bonds, $n_{+-}/(2L^2)$.
We are thus led to the same formula~(\ref{eq:1Dlt}) as for the linear chain,
\beq\label{eq:xiEdef}
\en(t)
=1-\frac{2}{\xi_{E}(t)},
\eeq
where $2/\xi_{E}(t)$ is the excess density of energy.

\subsection{Existence of blocked and long-lived metastable configurations}

\begin{figure}[ht]
\begin{center}
\includegraphics[angle=0,width=0.40\linewidth]{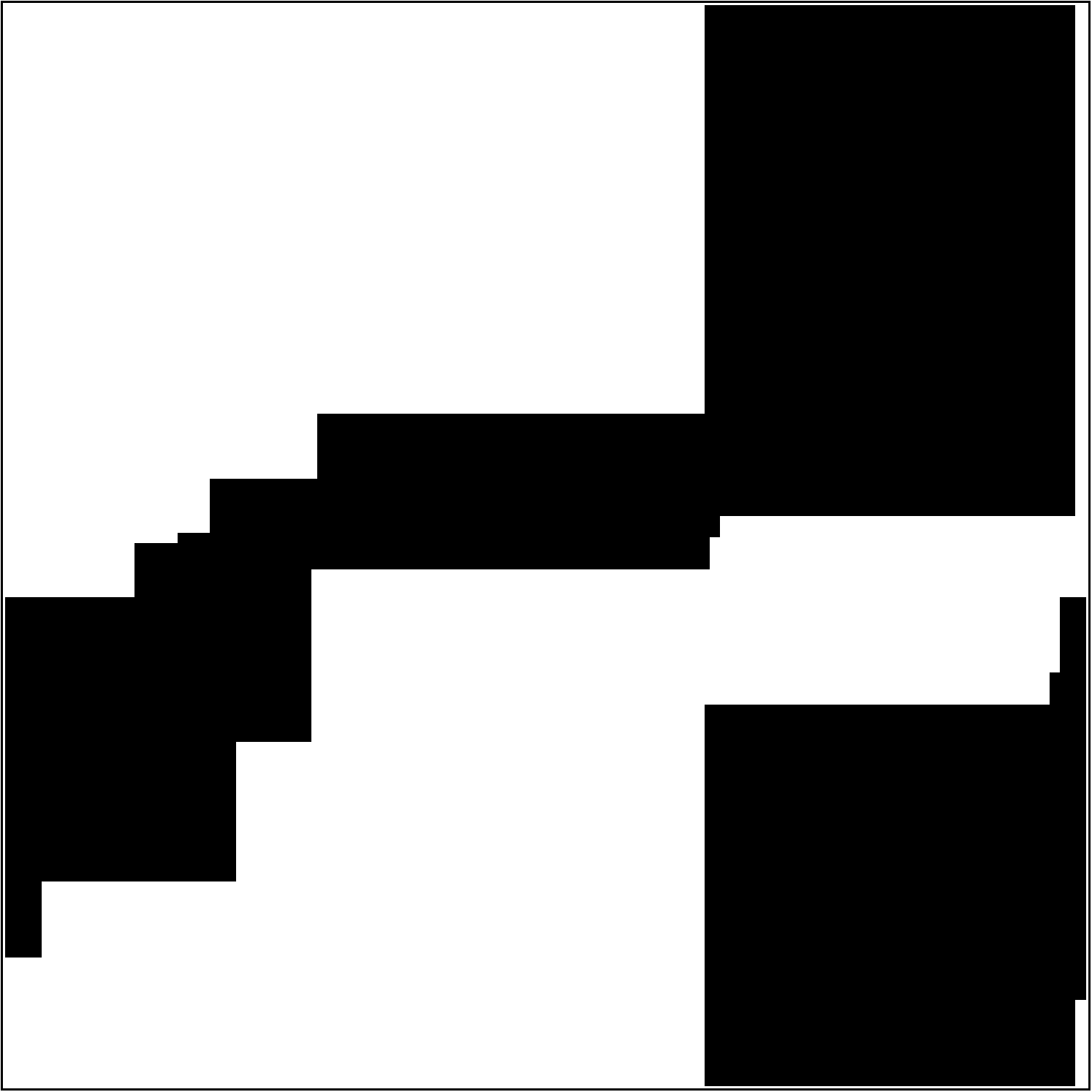}~~~~~~~
\includegraphics[angle=0,width=0.40\linewidth]{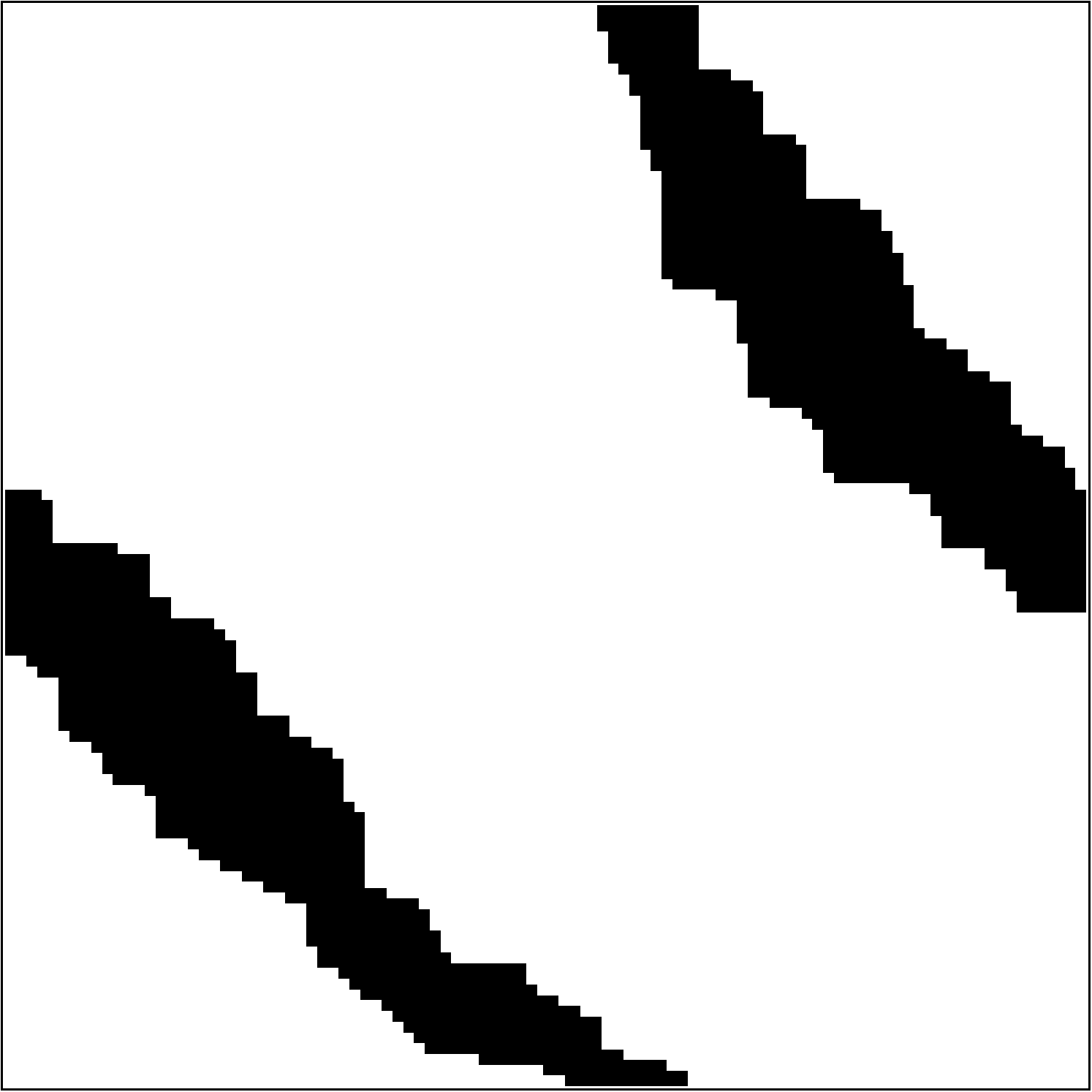}
\end{center}
\caption{
Left panel: snapshot of a blocked configuration with $\en=\en_2=0.96$ and $V=0$.
Right panel: long-lived configuration with $\en=\en_2=0.96$ and $V=1$ which eventually converges to the ground state. 
This is the configuration corresponding, in figure~\ref{fig:e}~(d), to one of the two rapidly evolving time trajectories from 0.96 to 1.
The system size is $L=100$.
}
\label{fig:snapp}
\end{figure}

\begin{figure}[ht]
\begin{center}
\includegraphics[angle=0,width=0.4\linewidth]{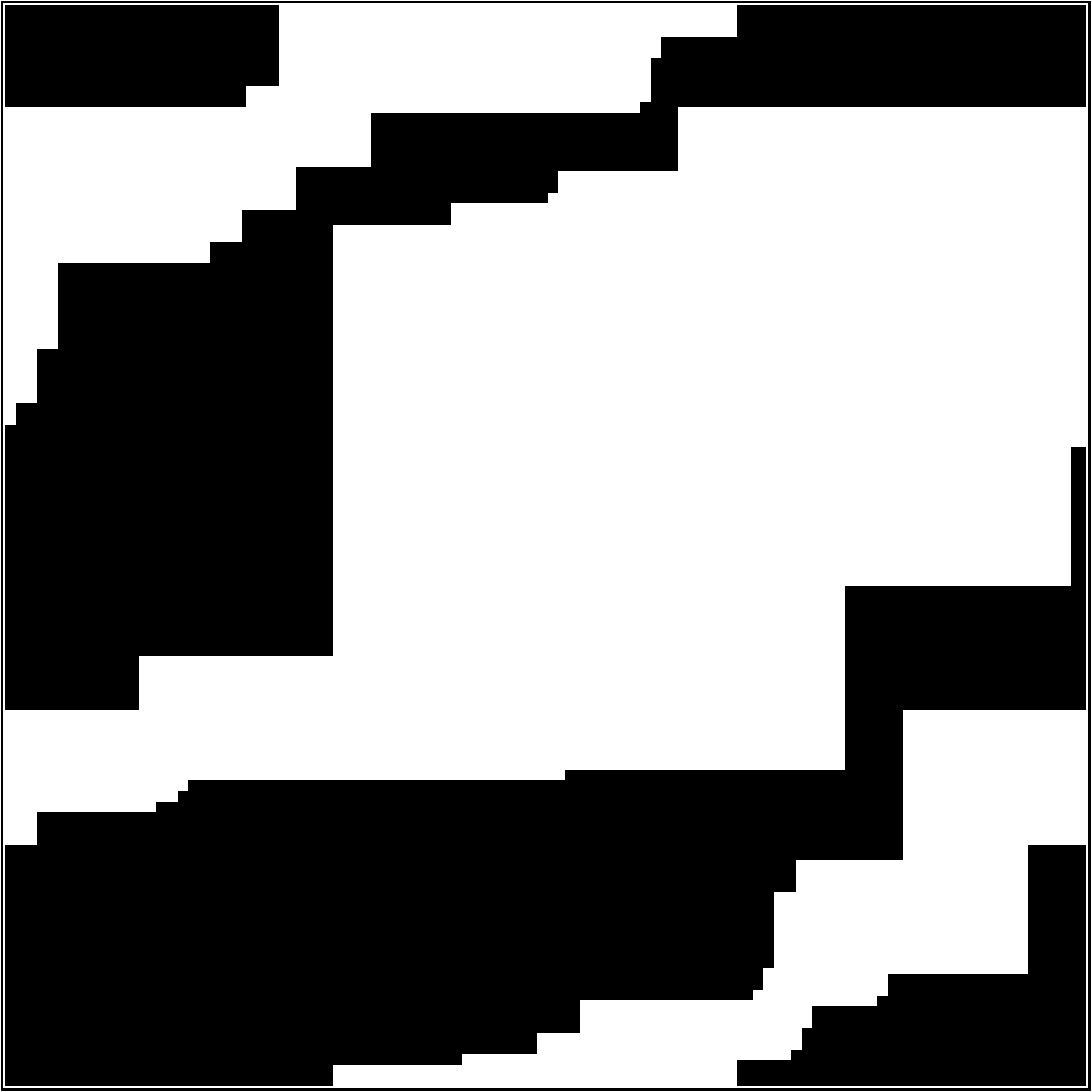}
\end{center}
\caption{
Snapshot of a (rare) blocked configuration with $\en=\en_3=0.94$ and $V=0$.
The system size is $L=100$.
}
\label{fig:snapp2}
\end{figure}

Figure~\ref{fig:e} depicts $\en(t)$ for 300 histories, for a system size $L=100$, with $V=0, 0.5, 1$.
From the observation of these figures one can draw the following conclusions.
Firstly, the speed of convergence of the trajectories to their limits depend on $V,$ a feature in line with the observations made for the disappearance of a minority cluster in section~\ref{sec:lifetime}.
Secondly, since the final value of $\en(t)$ should be equal to 1 in the ground state, we see that a fraction of the histories 
do not reach the ground state and therefore correspond to configurations of higher energy.
We can indeed identify several groups of time-trajectories.
The first group converges to $\en_0=1$, the ground state, the other groups to lower values of $\en$: $0.98, 0.96$.
These plateau values are in accordance with~(\ref{eq:complex}):
$\en_1=0.98$ corresponds to blocked configurations which are either horizontal or vertical stripes;
$\en_2=0.96$ corresponds either to blocked configurations which are stripes parallel to the velocity (see left panel of figure~\ref{fig:snapp} for a snapshot of such a configuration) or to slowly disappearing configurations (see right panel of figure~\ref{fig:snapp} for a snapshot of such a configuration), which are stripes perpendicular to the velocity.
Note that smaller values of $\en$ can actually be reached if the number of histories is larger.
For instance, for 4800 randomly selected histories, stripes with $\en=e_3 =0.94$ are encountered, see~\textcolor{black}{figure~\ref{fig:snapp2}} for an example.
They correspond to configurations with two stripes parallel to the velocity.
We also observe on figures~\ref{fig:e} that the group of histories ending in the ground state (in black) is the largest one, followed by those ending on $e_1$ (in red), then on $e_2$ (in green).
A last observation concerns the longer scale of time corresponding to the long-lived metastable states mentioned above.
This can be seen on figure (d) where after a long time two histories which seemingly had ended on $e_2=0.96$ finally jump to $e_0=1$.
Note that these `jumping' trajectories do exist for any value of $V$.

We relegate the quantitative study of the blocked configurations to another work.

\begin{figure}[ht]
\begin{center}
\includegraphics[angle=0,width=0.6\linewidth]{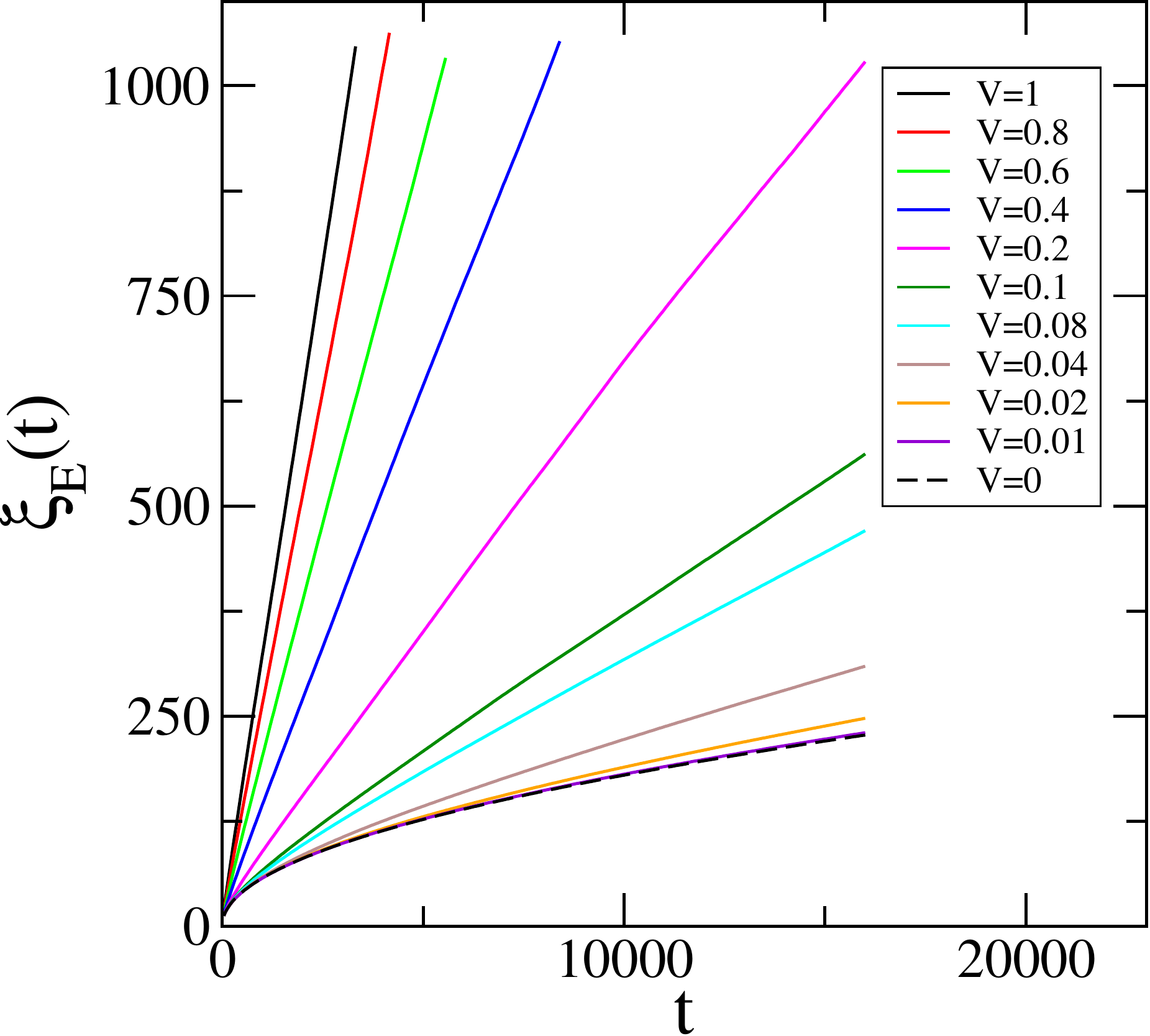}
\end{center}
\caption{
Typical growing length $\xi_{E}(t)$ for $V=0,0.01,\dots, 1$.
System sizes range from $L=1400$ for $V=0$ to $L=2000$ for $V=1$.
}
\label{fig:xiE}
\end{figure}
\begin{figure}[ht]
\begin{center}
\includegraphics[angle=0,width=0.4\linewidth]{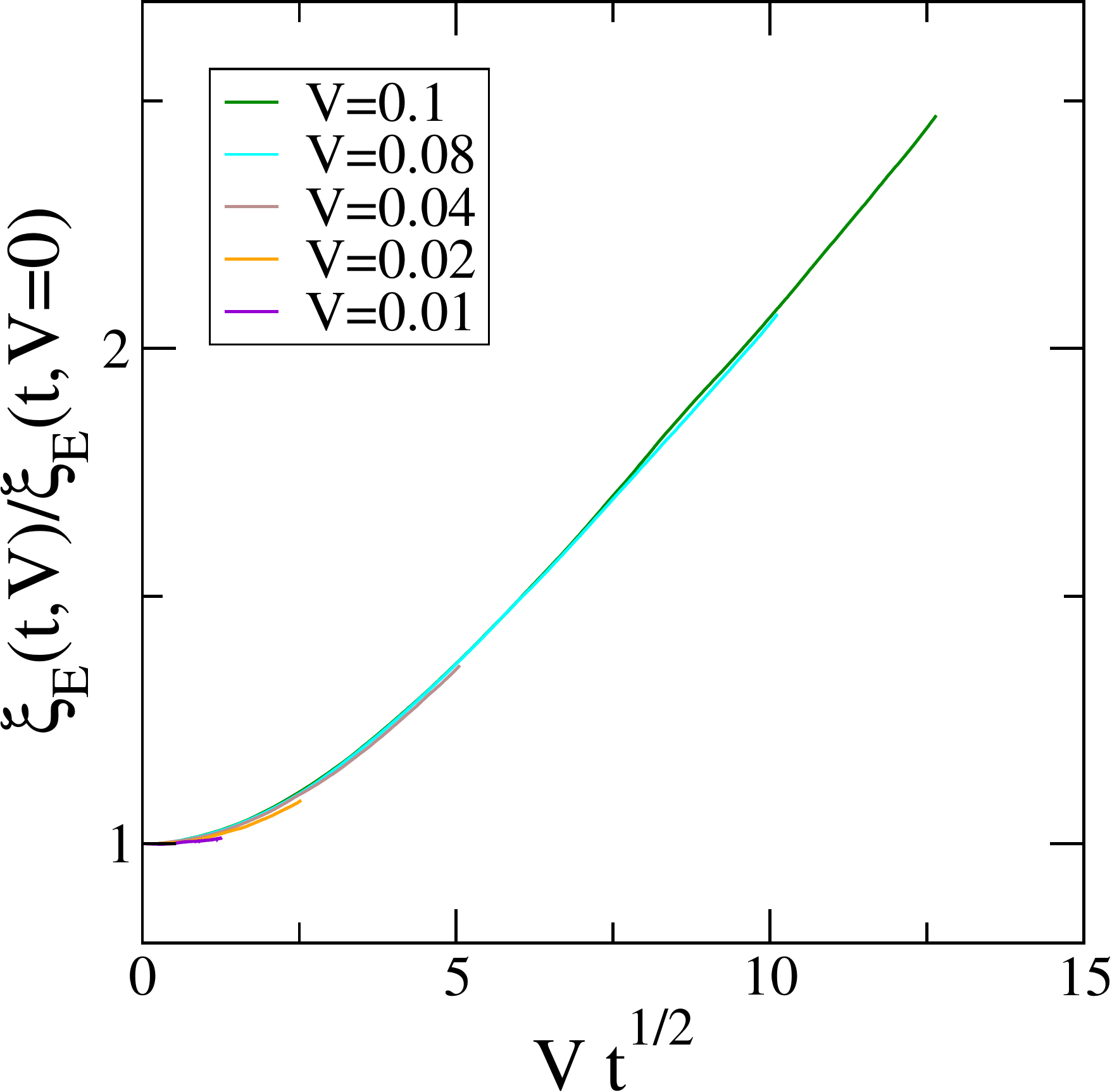}
\includegraphics[angle=0,width=0.4\linewidth]{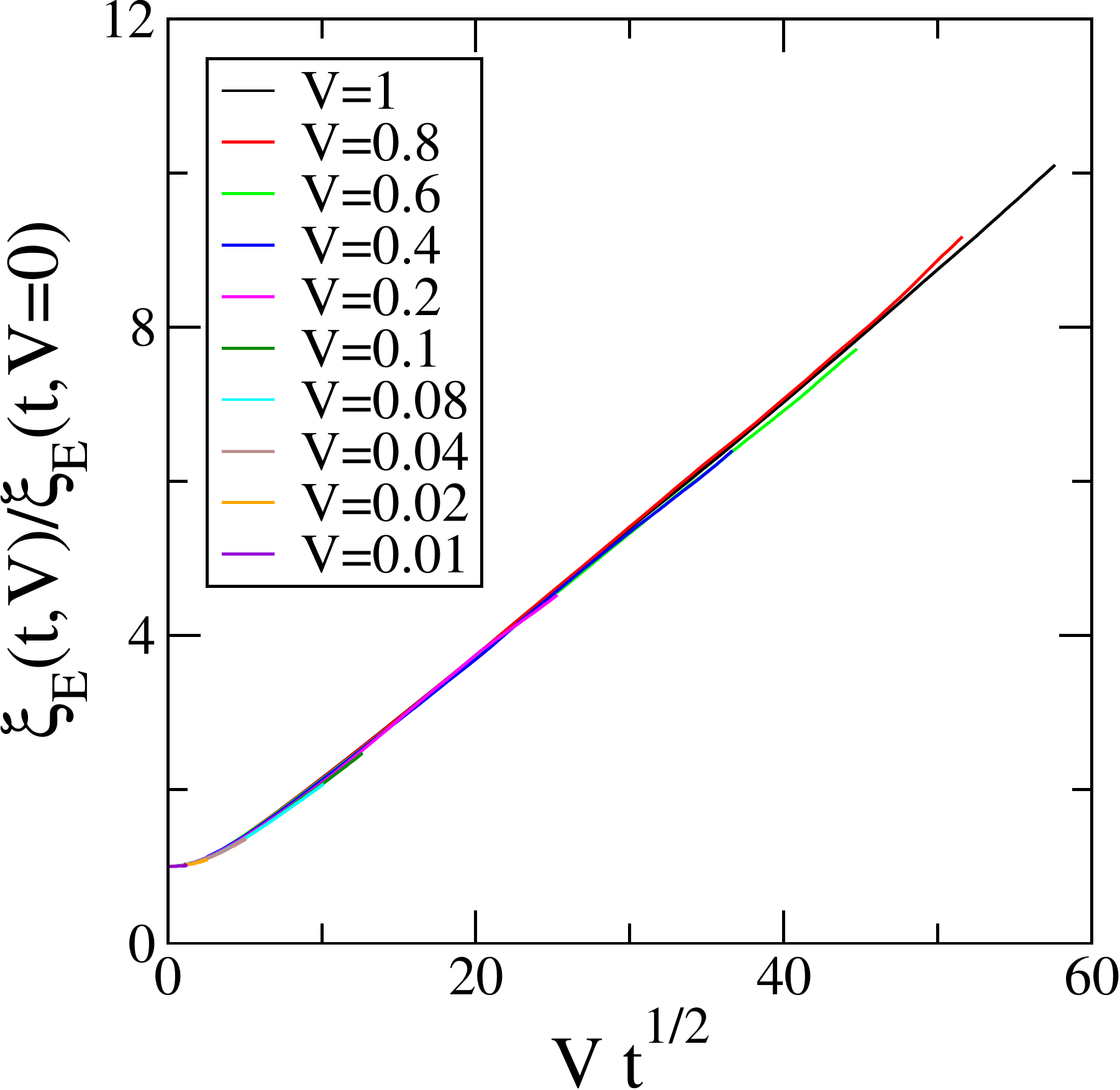}
\end{center}
\caption{Scaling function~(\ref{eq:xiEscal}) showing the crossover from diffusive coarsening to ballistic coarsening.
Same data as in figure~\ref{fig:xiE} for all range of values of $x=V\sqrt{t}$.
}
\label{fig:xiEscaling}
\end{figure}

%
\subsection{Scales of time for the coarsening regime defined by the growing length $\xi_{E}(t)$}

We now want to determine the scales of time of the (transient) coarsening regime, before $\en(t)$ (and therefore $\xi_{E}(t)$) reaches a plateau value.
Figure~\ref{fig:xiE} depicts $\xi_{E}(t)$ (see~(\ref{eq:xiEdef})) against $t$ for various values of $V$.
For $V=0$ we observe a power-law growth of $\xi_{E}(t)$ as $t^{1/2}$, before reaching the plateau value,
as for the well-studied Glauber case~\cite{bray,humayun}.
We then observe, for small values of $V$, a crossover to a linear dependence of $\xi_{E}(t)$ in $V$.

This crossover from diffusive coarsening to ballistic coarsening is best revealed by assuming the scaling form for the ratio
\beq\label{eq:xiEscal}
\frac{\xi_{E}(t,V)}{\xi_{E}(t,0)}\approx h_{E}(x=V\sqrt{t}),
\eeq
with obvious notations.
Figure~\ref{fig:xiEscaling} depicts the scaling function $h_{E}(x)$.
For $x$ very small $h_{E}(x)$ is constant, corresponding to the diffusive regime, then it crosses over quadratically to the ballistic regime, when $x$ increases.
At larger $x$ the scaling function becomes linear in $x$, i.e., $\xi_{E}(t)$ grows as $Vt$ and
this linear behaviour extends to even larger values of the scaling variable $x$, as demonstrated in figure~\ref{fig:xiEscaling} right panel.

\subsection{Scales of time for the coarsening regime defined by the time to reach a plateau}

We complement the study done above by measuring the characteristic times for the trajectories identified in figures~\ref{fig:e} to reach a plateau value.
We restrict our study to trajectories reaching the ground state ($\en=\en_0=1$) or the blocked state with $\en=e_1$.

We specialize first to trajectories reaching the ground state ($\en_0=1$).
As demonstrated by figure~\ref{fig:e} (d) these trajectories fall into two classes: those reaching the ground state directly and those reaching the ground state with a delay, because they first spend a long time in a long-lived configuration\footnote{See the right panel of figure~\ref{fig:snapp} which depicts one of the two long-lived configurations of figure~\ref{fig:e}~(d) with $\en_2=0.96$ ($L=100$) eventually reaching the ground state.}.
The great majority of the trajectories belong to the first class.
Therefore a simple way of defining a typical time avoiding the difficulty of discriminating between the two classes consists in recording the distribution of all times 
$\tau_0$ for reaching the ground state and then measure the median of this distribution, 
denoted by $\overline{\tau}_0$,
for a given maximal time of observation $t_{\rm max}$ (and a given number of histories).
Indeed the occurrence of trajectories which eventually converge to $\en_0=1$ at very late times does not affect the value of $\overline{\tau}_0$ while it would affect the value of the average $\langle \tau_0\rangle$.

\begin{figure}[ht]
\begin{center}
\includegraphics[angle=0,width=0.9\linewidth]{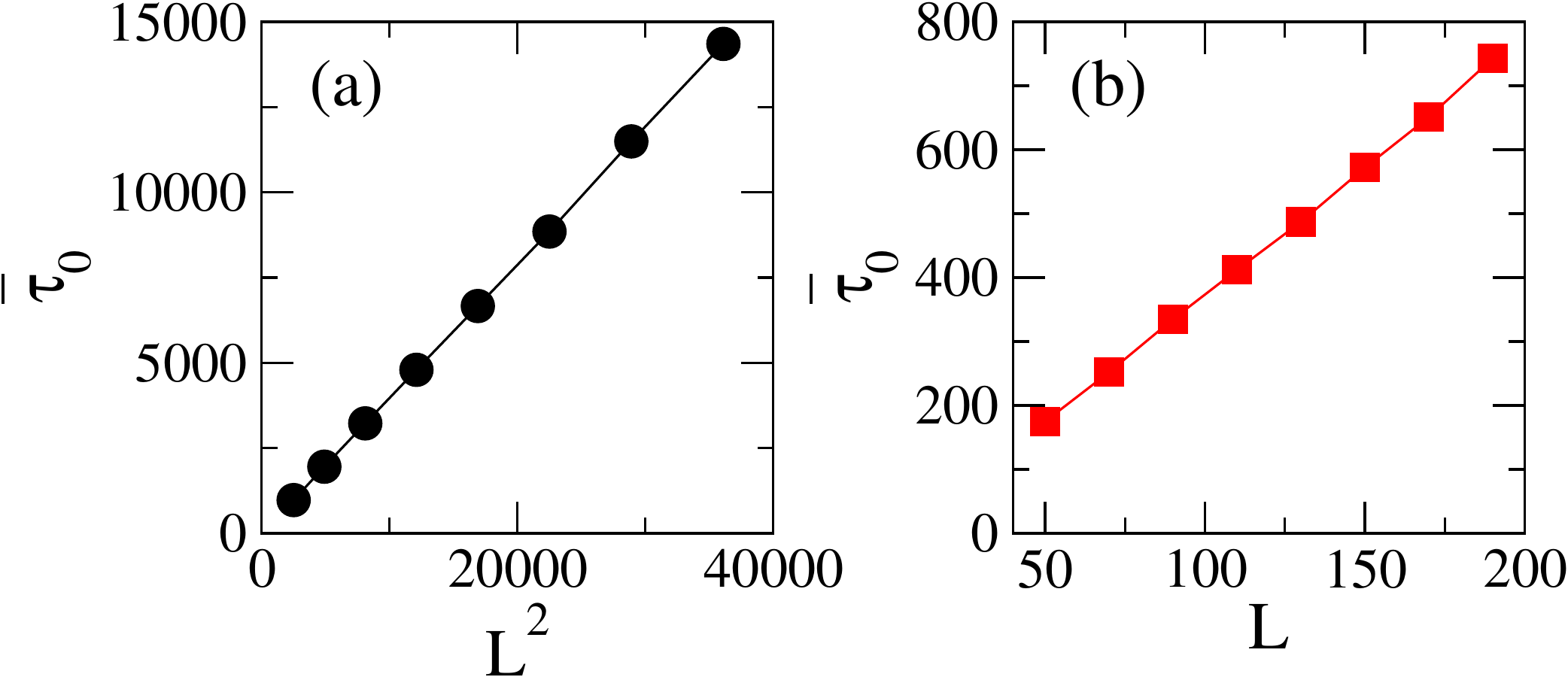}
\end{center}
\caption{
Median $\overline{\tau}_0$ of the distribution of times $\tau_0$ to reach the ground state, for a given maximal time of observation $t_{\rm max}=200\,000$.
For every system size 4000 histories are considered.
Plots are against $L^2$ for $V=0$ (figure (a)) and against $L$ for $V=1$ (figure (b)).
}
\label{fig:tau_gs}
\end{figure}

This protocol results in \textcolor{black}{figures~\ref{fig:tau_gs}}.
For $V=0$ this time is proportional to $L^2$, as it should (diffusive regime).
For $V=1$ this time is proportional to $L$: the dynamics of coarsening is accelerated, it crosses over from diffusive to ballistic.
We observe good convergence of $\overline{\tau}_0$ to a definite value as soon as $N_{{\rm history}}>1000$.

\begin{figure}[ht]
\begin{center}
\includegraphics[angle=0,width=0.6\linewidth]{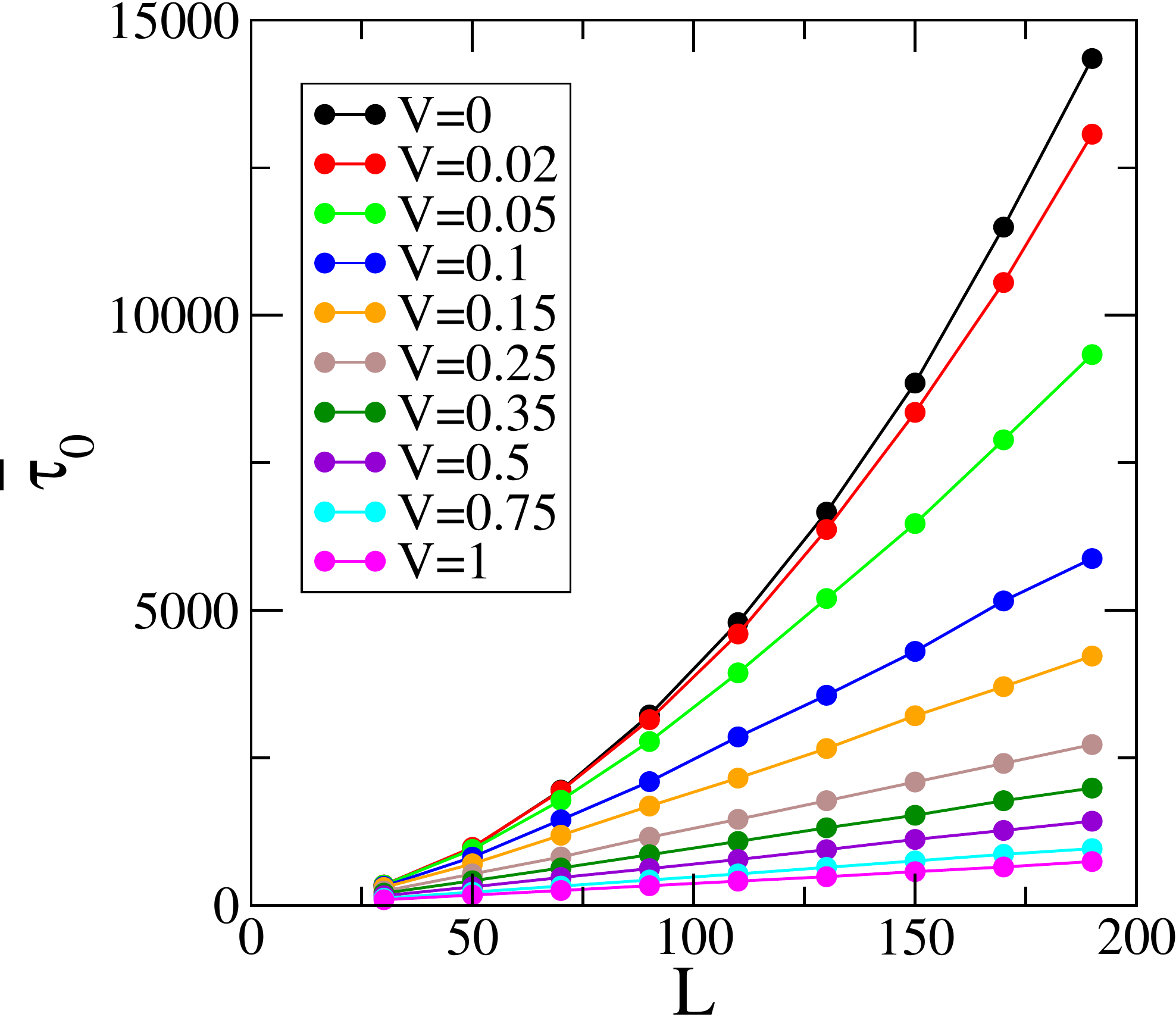}
\end{center}
\caption{Median $\overline{\tau}_0$ of the distribution of times $\tau_0$ to reach the ground state as a function of system size $L$,
for values of $V$ ranging from $0$ to $1$. 
The number of histories used for 
each value of $V$ is equal to 4000 and $t_{\rm max}=200\,000$.
}
\label{fig:tau}
\end{figure}

\begin{figure}[ht]
\begin{center}
\includegraphics[angle=0,width=0.6\linewidth]{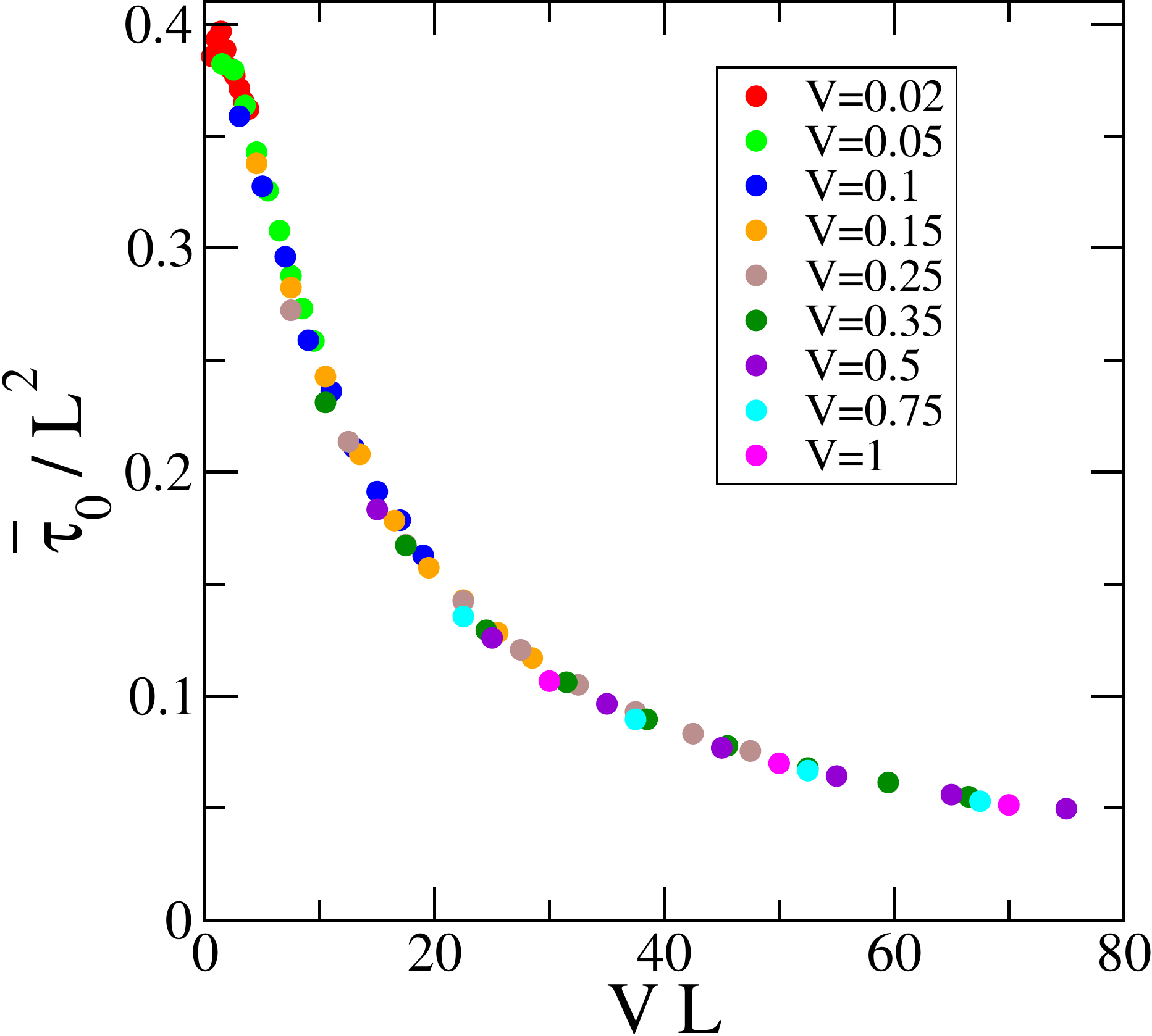}
\end{center}
\caption{Same data as in figure~\ref{fig:tau}, with scaling hypothesis~(\ref{eq:scal}).
}
\label{fig:tau_scal}
\end{figure}

We now measure $\overline{\tau}_0$ for various values of $V$ and $L$.
We find the curves of figure~\ref{fig:tau}.
We analyze the data by plotting them with the scaling assumption 
\beq\label{eq:scal}
\overline{\tau}_0=L^2 g_\tau(V L).
\eeq
We predict that for $x=VL$ small, $g_{\tau}(0)$ is finite, hence $\overline{\tau}_0$ scales as $L^2$;
for $x$ large, $g_{\tau}(x)\sim 1/x$ hence $\overline{\tau}_0$ scales as $VL$.
The scaling function $g_{\tau}(x)$ describing the crossover from diffusive to ballistic is depicted in figure~\ref{fig:tau_scal}.

\begin{figure}[ht]
\begin{center}
\includegraphics[angle=0,width=0.9\linewidth]{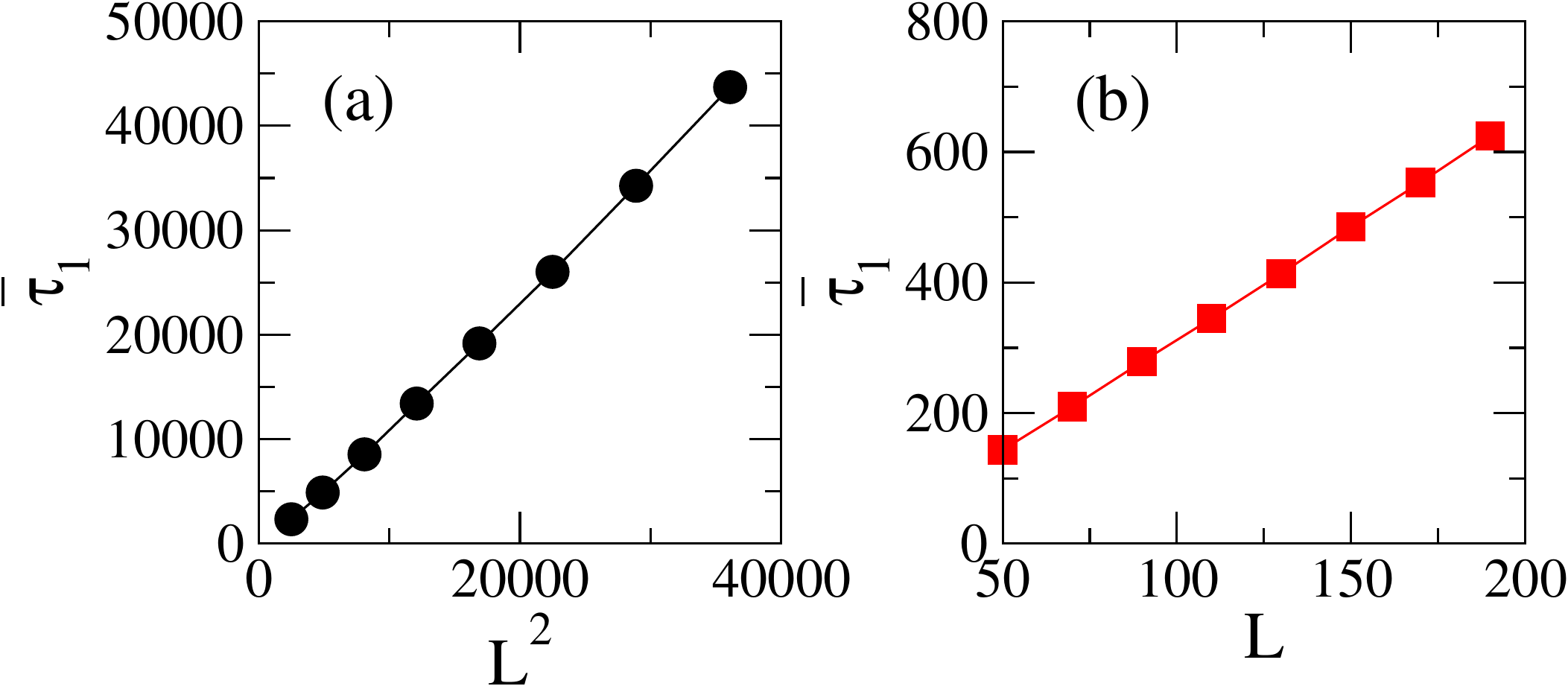}
\end{center}
\caption{
Median $\overline{\tau_1}$ of the distribution of times $\tau_1$ to reach the blocked state $e_1$, for a given maximal time of observation $t_{\rm max}=200\,000$.
For every system size 4000 histories are considered.
Plots are against $L^2$ for $V=0$ (figure (a)) and against $L$ for $V=1$ (figure (b)).
}
\label{fig:tau_k2_gs}
\end{figure}

The same analysis can be performed for the time $\tau_1$ to reach the blocked state $e_1$.
Figure~\ref{fig:tau_k2_gs} depicts the median $\overline{\tau_1}$ of the distribution of times $\tau_1$, for a given maximal time of observation $t_{\rm max}=200\,000$.
Plots are against $L^2$ for $V=0$ and against $L$ for $V=1$.
The same conclusions hold.

To summarize, as demonstrated in this section, coarsening is diffusive for reversible dynamics with $V=0$, as for Glauber dynamics, while it becomes ballistic for irreversible dynamics with $V\ne0$.

\section{Equal-time correlation function}
\label{sec:Lt}

The aim of this section is to complement the study made above by the investigation of the equal-time correlation function.
This allows both to confirm the crossover from diffusive to ballistic coarsening as soon as $V\ne0$ and to demonstrate the existence of an anisotropy in the dynamics of our model.

\subsection{Definition of $\xi_{C}(t)$}
\label{sec:defLt}

Let us consider the equal-time correlation function
\beqa \label{eq:crt}
C(r,t) 
&=&\langle \s_{0,0}(t) \s_{i,j}(t) \rangle,
\eeqa
where $r^2=i^2+j^2$. 
In the following we consider this correlation in different directions. 
The directions parallel to the axes, with either $i =0$ or $j=0$, will be denoted by $(1,0)$. 
The North-East and North-West directions, where $j=i$ or $j=-i$, respectively,
will be denoted by $(1,1)$ and $(1,-1)$.

\begin{figure}[ht]
\begin{center}
\includegraphics[angle=0,width=0.6\linewidth]{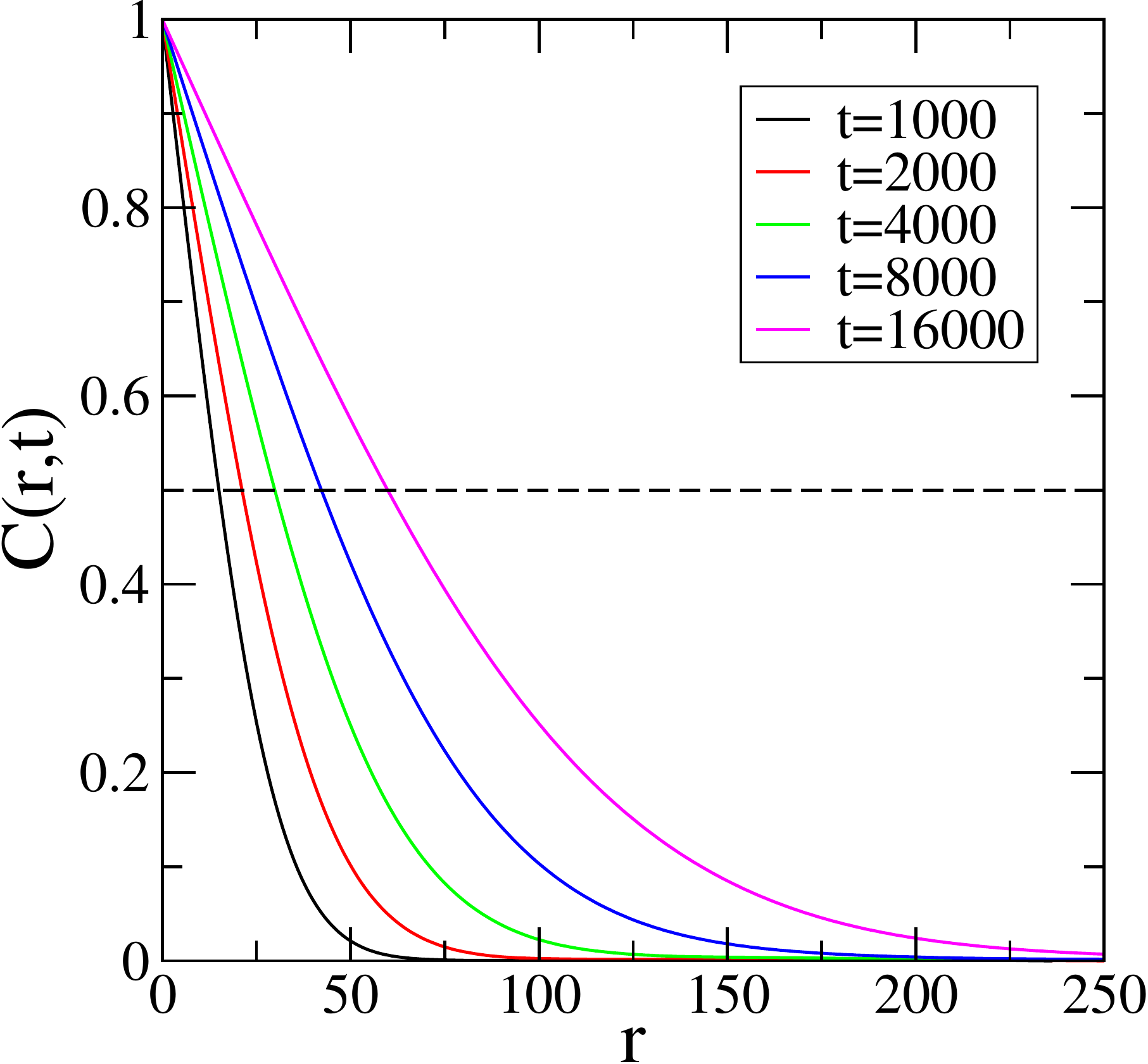}
\end{center}
\caption{Protocol to extract $\xi_{C}(t)$ from $C(r,t)$ ($V=0, L=1400$).
}
\label{fig:CrtProtoc}
\end{figure}

In the coarsening regime, the correlation function (\ref{eq:crt}) allows to define an alternate growing length, denoted by $\xi_{C}(t)$, as follows.
We determine the intersections of $C(r,t)$ with a 
constant line $C=C_0$. 
The value used in the following
is $C_0 = 0.5$, i.e., $\xi_{C}(t)$ is obtained as the half-height width of $C(r,t)$ (see figure~\ref{fig:CrtProtoc}), but we checked that other values yield the same results (up to a pre-factor) for the growing length $\xi_{C}(t)$.

\begin{figure}[ht]
\begin{center}
\includegraphics[angle=0,width=0.6\linewidth]{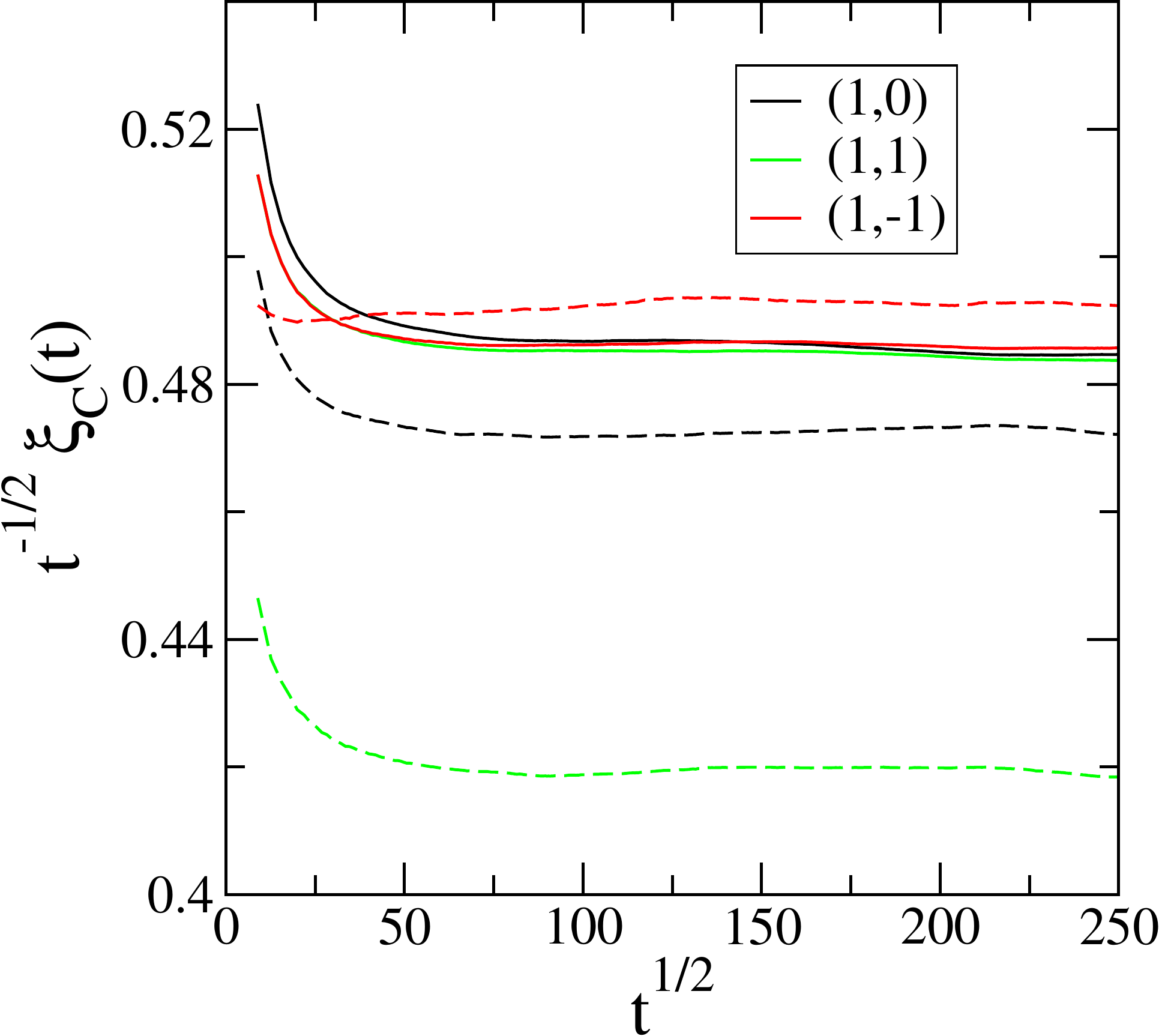}
\end{center}
\caption{Three different lengths extracted from the equal-time correlation $C(r,t)$ in three different directions
for Glauber dynamics (full lines) and $V=0$ dynamics (dashed lines). 
The data result from averaging over at least 1000 independent histories. 
The system size is $L=1400$.
}
\label{fig_L_ani_1}
\end{figure}

\begin{figure}[ht]
\begin{center}
\includegraphics[angle=0,width=0.6\linewidth]{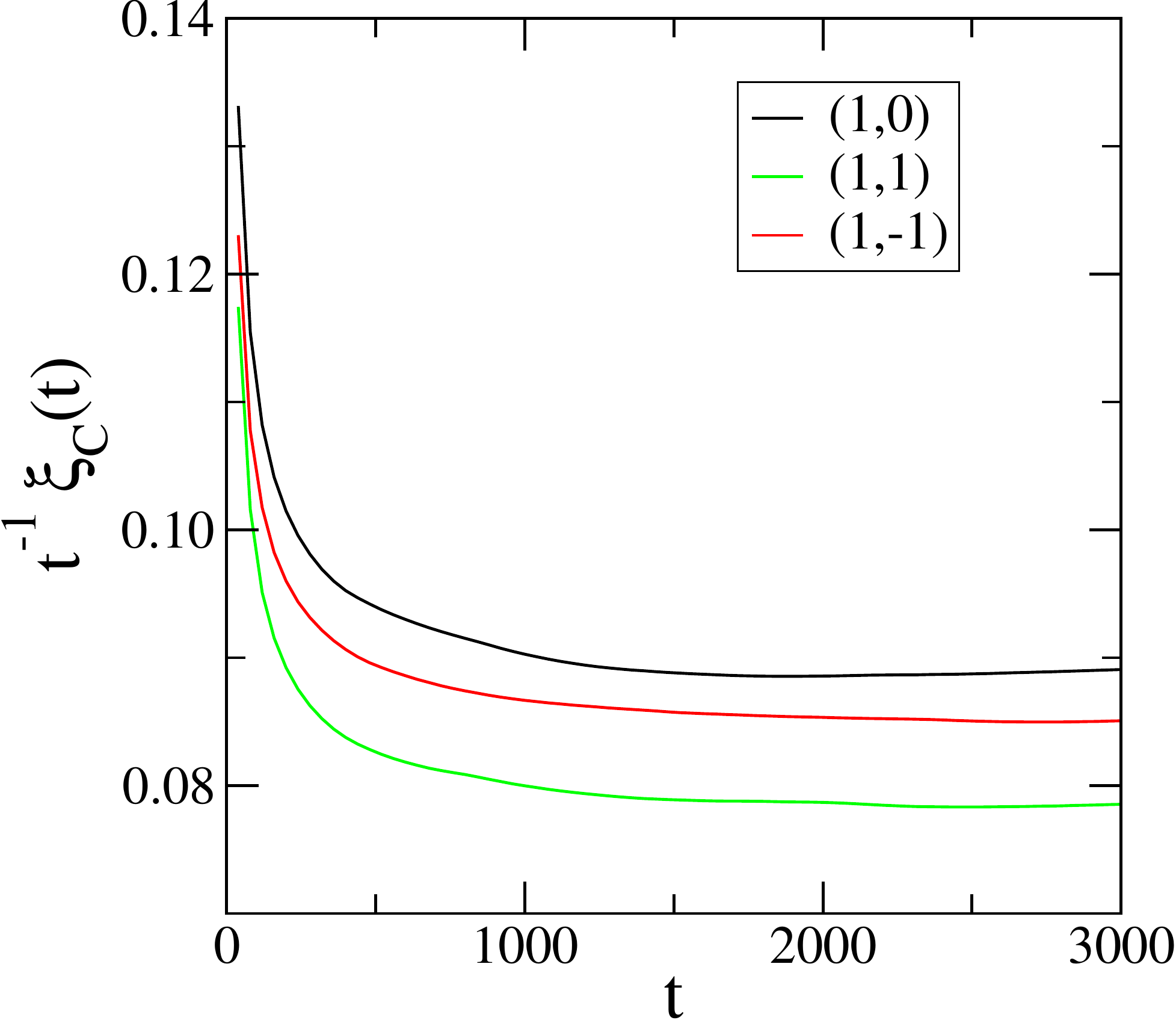}
\end{center}
\caption{Three different lengths extracted from the equal-time correlation $C(r,t)$ in three different directions
for $V=1$ dynamics. 
The data result from averaging over at least 1000 independent histories. 
The system size is $L=2000$.
}
\label{fig_L_ani_2}
\end{figure}

We measured the three different lengths $\xi_{C}(t)$ corresponding respectively to the $(1,0)$, $(1,1)$ and $(1,-1)$ directions, both for Glauber and $V=0$ dynamics.
For both dynamics all three lengths increase as $t^{1/2}$ as shown in figure~\ref{fig_L_ani_1}.
Whereas no anisotropy shows up for Glauber dynamics, a direction-dependent pre-factor modifies the lengths
for $V=0$ dynamics. 
A similar anisotropy is encountered when $V \ne 0$, as seen on figure~\ref{fig_L_ani_2}.
In the following we restrict to the (1,0) direction for the analysis.

\begin{figure}[ht]
\begin{center}
\includegraphics[angle=0,width=0.6\linewidth]{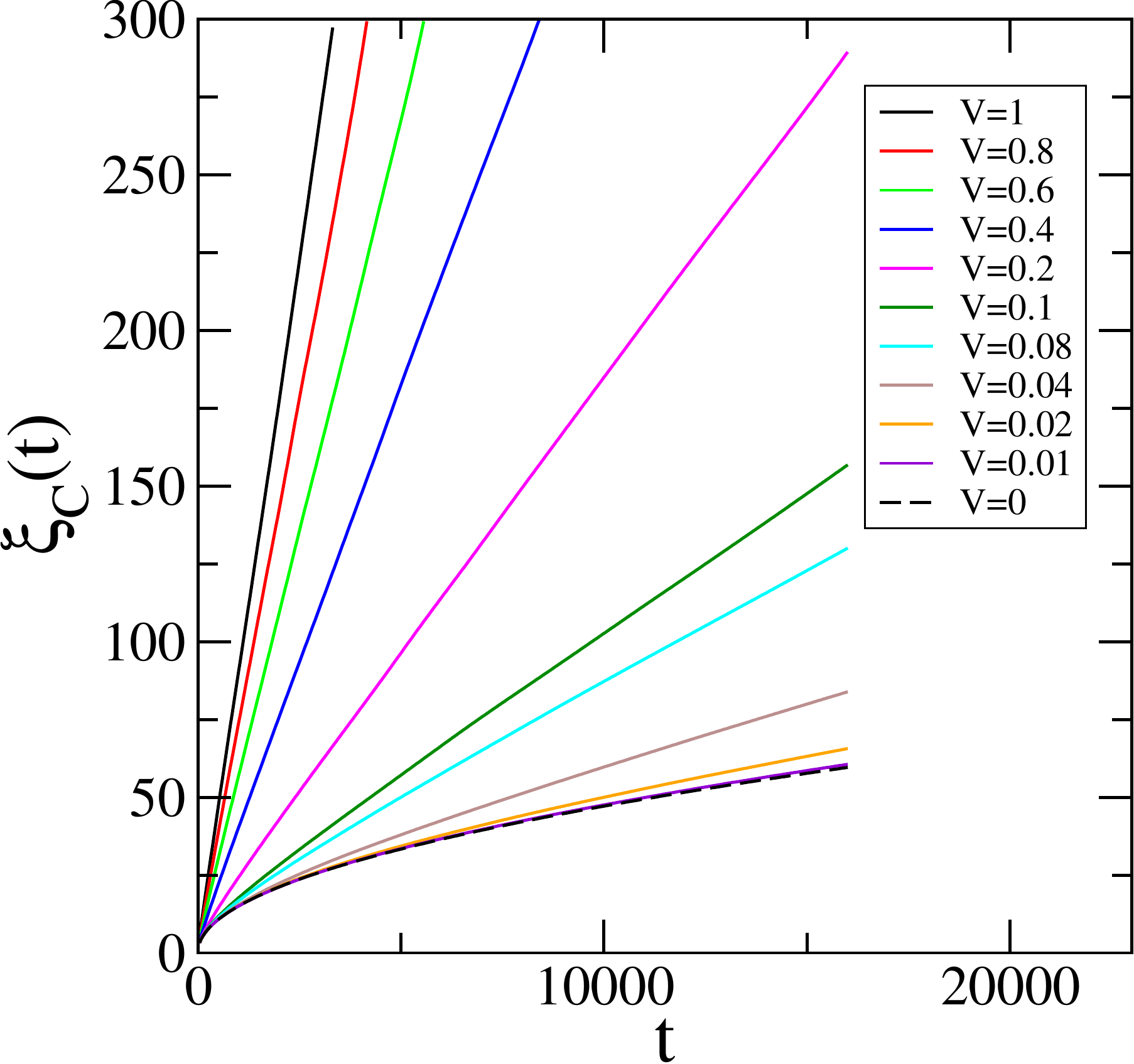}
\end{center}
\caption{Typical growing length $\xi_{C}(t)$ for $V=0,0.01,\dots, 1$.
System sizes range from $L=1400$ for $V=0$ to $L=2000$ for $V=1$.
At least 1000 histories have been used for the averages.
}
\label{fig:xiC}
\end{figure}

\begin{figure}[ht] 
\begin{center}
\includegraphics[angle=0,width=0.4\linewidth]{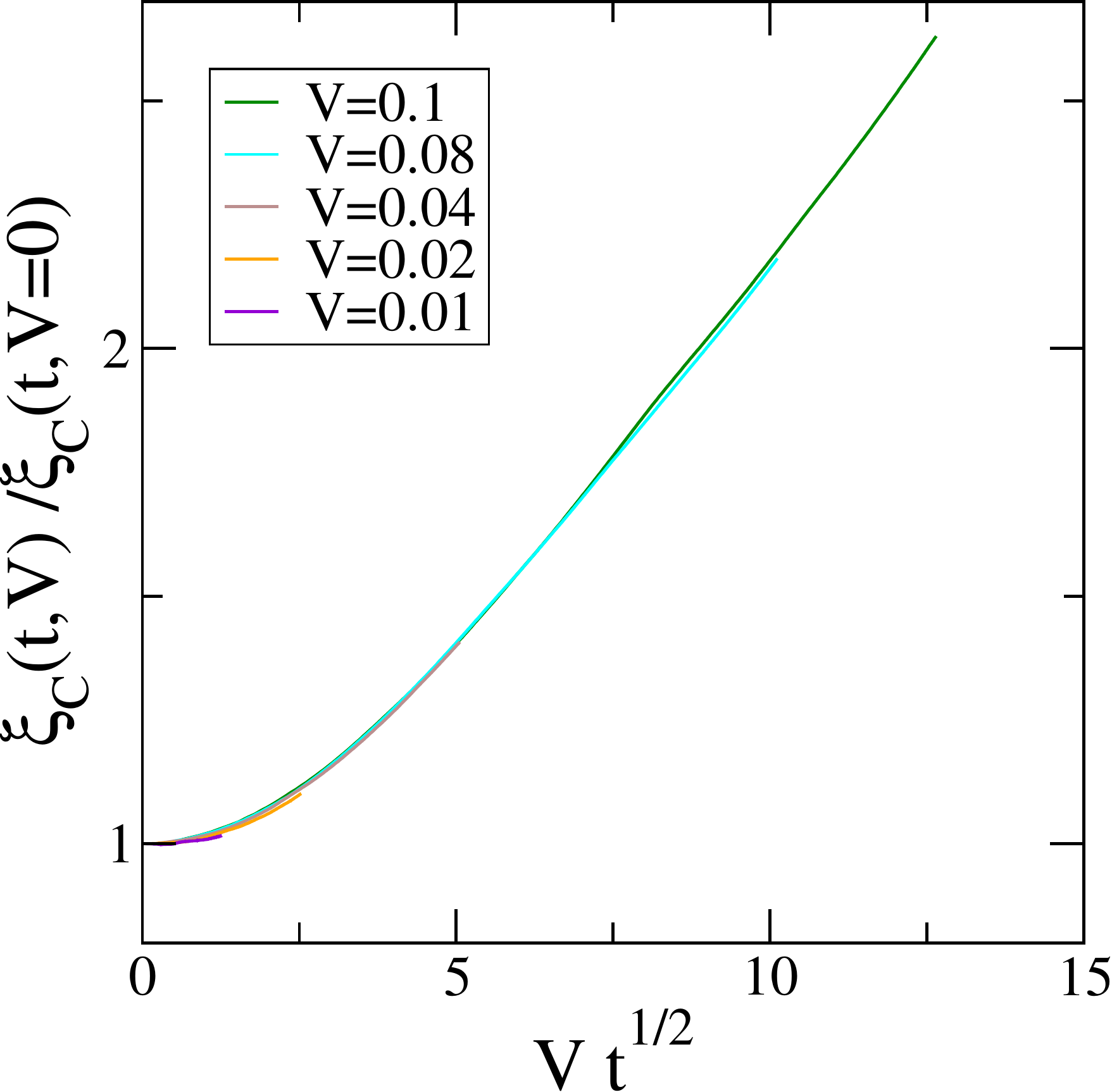}
\includegraphics[angle=0,width=0.4\linewidth]{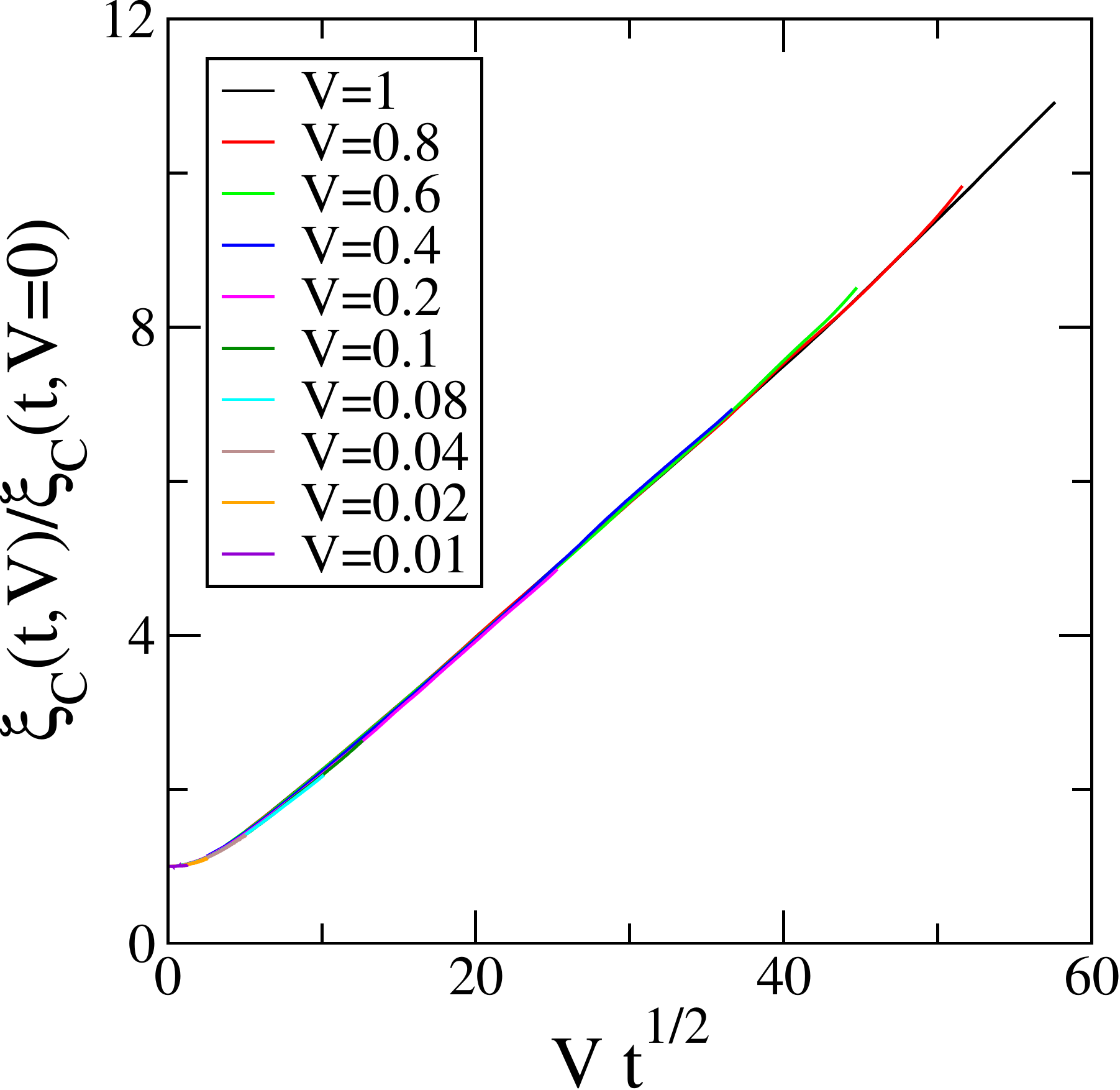}
\end{center}
\caption{Left panel: scaling function~(\ref{eq:xiCscal}) showing the crossover from diffusive coarsening to ballistic coarsening.
Right panel: same data as in figure~\ref{fig:xiC} for all range of values of $x=V\sqrt{t}$.
}
\label{fig:xiCscal}
\end{figure}

Figure~\ref{fig:xiC} depicts $\xi_{C}(t)$ for various values of $V$.
The crossover to the diffusive regime can be seen for small values of $V$.
As for $\xi_{E}(t)$, the crossover is best revealed by assuming the scaling form for the ratio
\beq\label{eq:xiCscal}
\frac{\xi_{C}(t,V)}{\xi_{C}(t,0)}\approx h_{C}(V\sqrt{t}).
\eeq
This scaling behaviour is demonstrated in the left panel of figure~\ref{fig:xiCscal}, which is, up to a proportionality constant, identical to the left panel of figure~\ref{fig:xiEscaling}.
Likewise, at large $x$ the scaling function becomes linear in $x$, i.e., $\xi_{C}(t)$ grows as $Vt$.
This linear behaviour extends to larger values of the scaling variable $x$, as demonstrated in the right panel of figure~\ref{fig:xiCscal}. 

\begin{figure}[ht]
\begin{center}
\includegraphics[angle=0,width=0.6\linewidth]{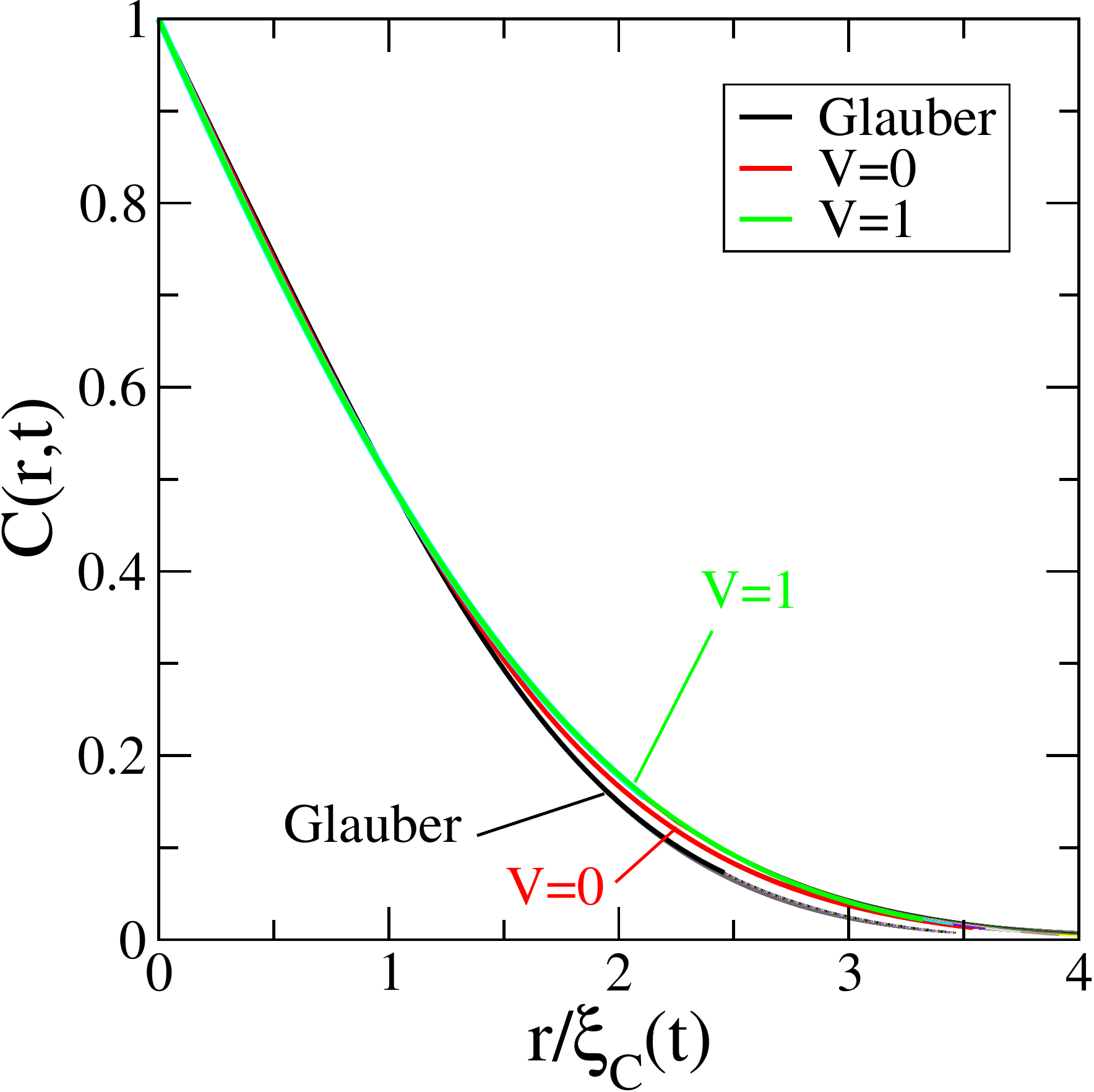}
\end{center}
\caption{Equal-time correlation function $C(r,t)$ as a function of $r/\xi_{C}(t)$.
For Glauber and $V=0$ dynamics the system size is $L=1400$, whereas for $V=1$ systems of length $L=2000$ have
been simulated. 
The data result from averaging over at least 1000 independent histories.
}
\label{fig:C_scal}
\end{figure}

We finally investigate the dependence of $C(r,t)$ both in space and time.
For Glauber dynamics and more generally for reversible non-conserving dynamics, it is known that the correlation function (\ref{eq:crt}) depends on space and time only through the
scaling variable $r/\xi_{C}(t)$~\cite{bray,humayun}:
\beq
C(r,t)=g_{C}\left(\frac{r}{\xi_{C}(t)}\right).
\eeq
As shown in figure~\ref{fig:C_scal} 
this also holds for $V=0$ and $V=1$. 
The three scaling functions are slightly different. 
Note that by the very definition of $C(r,t)$ the three curves are constrained to intersect at the point $(1,1/2)$.

We finally compared $\xi_{E}(t)$ to $\xi_{C}(t)$.
These two lengths, which are two alternate representations of the same reality, appear to be essentially proportional to each other,
as can be seen on figures~\ref{fig:xiE} and~\ref{fig:xiC}, with a constant of proportionality slightly dependent on $V$.

\begin{figure}[ht]
\begin{center}
\includegraphics[angle=0,width=0.6\linewidth]{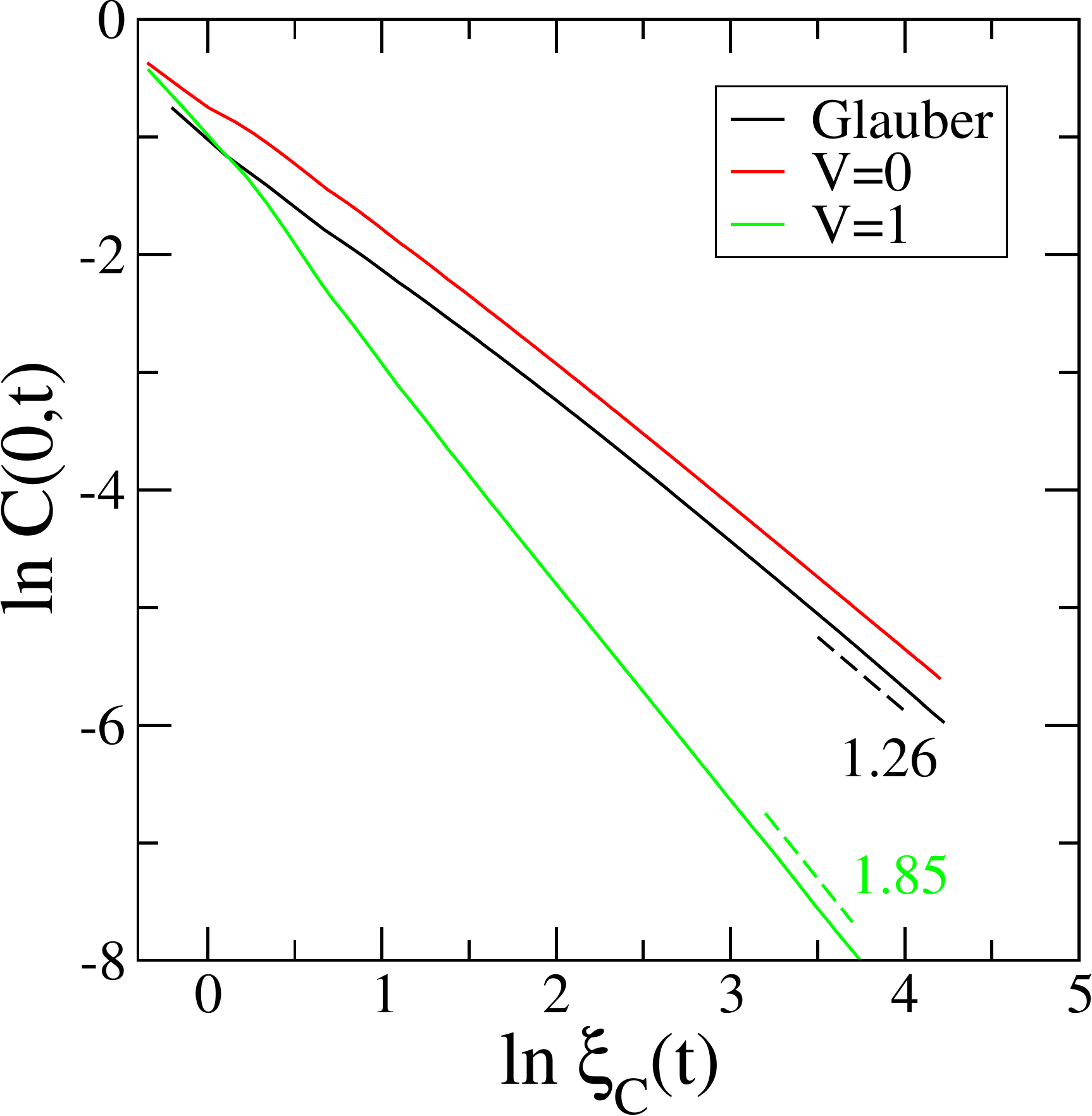}
\end{center}
\caption{Autocorrelation plotted against $\xi_{C}(t)$ for Glauber dynamics as well as for $V=0$ and $V=1$ dynamics.
The slopes in the long-time regime give the autocorrelation exponent $\lambda$.
For Glauber and $V = 0$ dynamics, $80\,000$ systems
with $L = 1200$ have been simulated. 
For $V = 1$ a total of $100\,000 $ systems with
$L = 2000$ were simulated.
}
\label{fig:autocor2}
\end{figure}

\section{Autocorrelation and persistence}

We finally proceed to the study of the autocorrelation function and of the probability of persistence. 
We also study the statistics of the mean temporal magnetization which gives another viewpoint on persistence.
The investigation of these quantities complement the study of the previous sections, which were only concerned by one-time observables.
The autocorrelation function is a two-time quantity while the persistence probability or the mean magnetization are multi-time objects, probing the entire history of the system.

Whether power-laws should survive or not in the ballistic regime is not obvious to predict a priori.
It turns out that both autocorrelation function and probability of persistence exhibit power-law decay in this regime,
with autocorrelation and persistence exponents larger than their counterparts for reversible dynamics.
The very existence of power-law behaviours is a confirmation of the existence of self-similar coarsening, even in the presence of irreversible dynamics (see also figure~\ref{fig:snapcoars} in section~\ref{sec:discussion}).

\subsection{Autocorrelation function}

For the two-dimensional Ising model quenched from infinite temperature down to zero temperature the autocorrelation
\beq
C(0,t)=\langle\s(0)\s(t)\rangle
\eeq
has a power-law decay at large times,
\beq
C(0,t)\sim \big(\xi_{C}(t)\big)^{-\lambda},
\eeq
with autocorrelation exponent $\lambda\approx 1.26$~\cite{autocorr,lambda}.
Figure~\ref{fig:autocor2} shows the results of simulations for the autocorrelation plotted against $\xi_{C}(t)$.
The results obtained for Glauber and $V=0$ dynamics are compatible with a common value $\lambda \approx 1.26$ for the autocorrelation exponent.
For the dynamics with $V=1$ the value found for the exponent, $\lambda \approx 1.85$, is markedly larger than for reversible dynamics, showing that the system decorrelates more rapidly under ballistic coarsening, as can be intuitively understood.

\subsection{Persistence}
\label{sec:persist}

For an Ising spin system evolving after a quench from high temperature down to zero temperature under non-conserved dynamics, the probability that a given spin did not flip up to time $t$ defines the persistence probability~\cite{dbg,bdg}.
This probability, denoted by $p(t)$, has a power-law decay at large time, with the persistence exponent $\theta$~\cite{dbg,bdg},
\beq
p(t)\sim t^{-\theta}.
\eeq
An accurate numerical analysis of this exponent for a two-dimensional system of Ising spins has recently been performed in~\cite{picco2}, which also reviews the existing literature on the subject.

Here our aim is to investigate, first, whether the anisotropy present in the dynamical rules of the $V=0$ dynamics changes the exponent $\theta$, 
and, secondly, whether under ballistic coarsening the persistence probability keeps a power-law decay, and if so with what value of the exponent.

\begin{figure}[ht]
\begin{center}
\includegraphics[angle=0,width=0.6\linewidth]{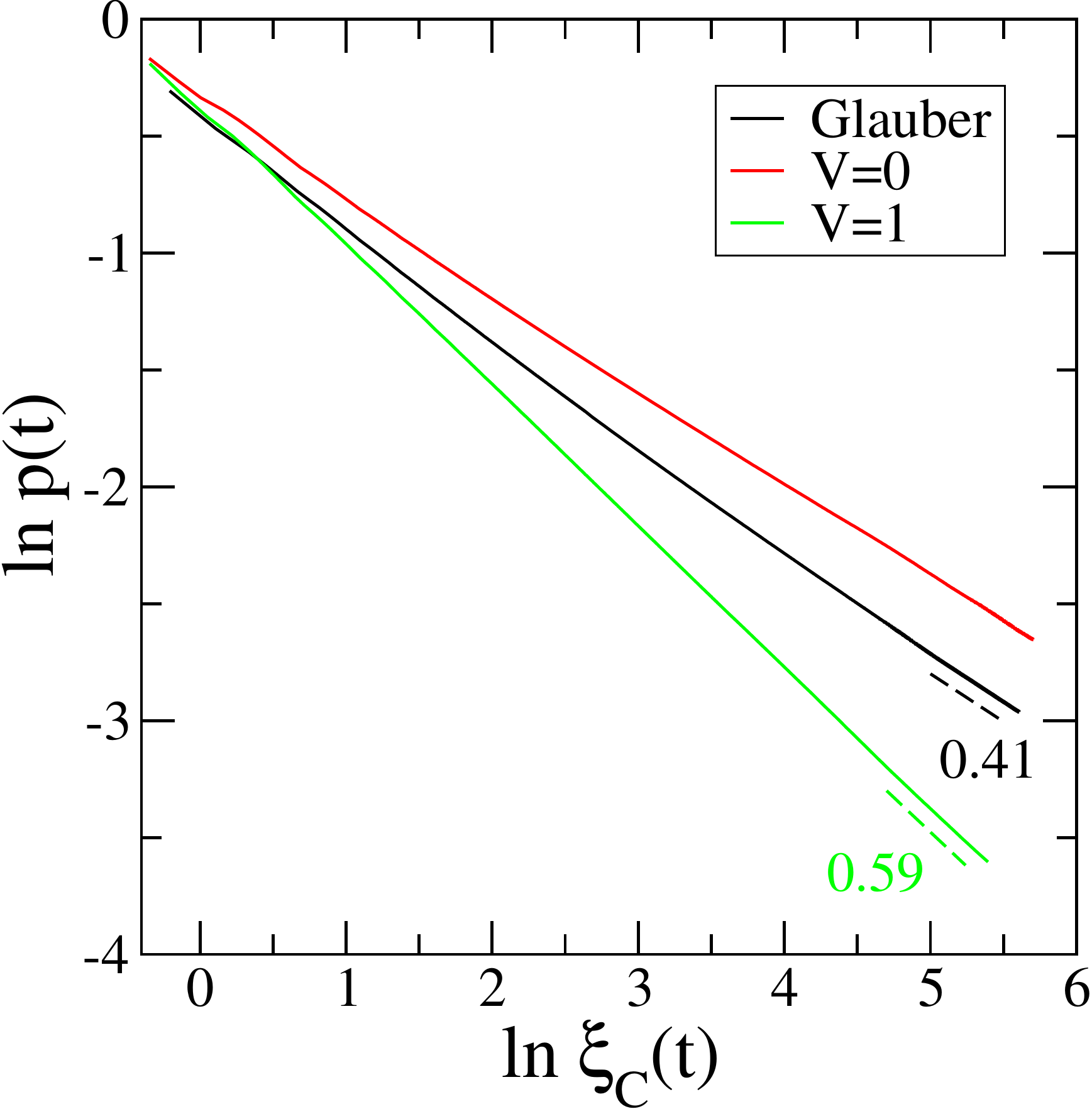}
\end{center}
\caption{Probability of persistence plotted against $\xi_{C}(t)$ for Glauber dynamics as well as for $V=0$ and $V=1$ dynamics.
The slopes 
yield estimates for $\theta z$ with $z=2$ for Glauber and $V=0$, whereas $z=1$ for $V=1$.
For Glauber and $V = 0$ dynamics the data, which have been obtained for systems of linear size $L=1200$, result from averaging over 200 independent histories.
For $V = 1$ a total of 320 systems with $L=2000$ were simulated. 
}
\label{fig:pers2}
\end{figure}

Figure~\ref{fig:pers2} depicts the persistence probability $p(t)$ against $\xi_C(t)$.
The behaviour for the $V=0$ dynamics seems to be the same as for Glauber dynamics, but it is difficult to conclude on the sole basis of this figure.
The common value of the persistence exponent has apparent value $\theta\approx 0.205$.
For $V=1$, a much larger value of the slope is obtained, yielding $\theta \approx 0.59$. 

Related to the concept of persistence is that of mean magnetization.
For a system of Ising spins the mean magnetization $M_t$ in the time interval $(0, t)$ is defined as~\cite{dg1998} 
\beq
M_t = \frac{1}{t} \int\limits_0^t \d t' \s(t'),
\eeq
where $\sigma(t')$ is the value of a selected spin at time $t'$.
The integral in the right side of this definition is equal to the difference $T_t^{+}-T_t^{-}$, where $T_t^{\pm}$ are the occupation times of the selected spin i.e., the times that this spin spends respectively in the $(+)$ or $(-)$ states up to time $t$.
These quantities therefore encode the statistics of flips that a given spin experiences up to time $t$.
The probability density of $M_t$ reads 
\beq
f_M(m;t)=-\frac{\d }{\d m}\prob(M_t\ge m).
\eeq
For a persistent spin, i.e.,
a spin which never flipped up to time $t$,
the value taken by $M_t$ is $m=\pm1$, depending on whether the initial value of this spin was $+1$ or $-1$,
and $\prob(M_t=\pm1)$ is just the persistence probability.
Hence the density $f_M(m;t)$ has discrete components at $m=\pm1$, equal to $p(t)/2$.

It was shown in~\cite{dg1998,drouffe1,drouffe2} that, in  the long-time limit, the distribution of $M_t$ has a limiting form $f_M(m)$, which is a U-shaped function with a singularity at $m=\pm1$ simply related to the persistence exponent $\theta$ 
as\footnote{The distribution of the mean magnetization can also be studied for the sign of a diffusion field, see~\cite{dg1998,tim1,tim2}. 
For Ising spin systems, extensions can be found in~\cite{drouffe1,drouffe2,balda}.}
\beq
f_M(m)\sim (1\mp m)^{\theta-1},\qquad(m\to\pm1).
\eeq
Thus this limiting process provides a stationary definition of persistence, even at finite temperature~\cite{drouffe1,drouffe2}.

\begin{figure}[ht]
\begin{center}
\includegraphics[angle=0,width=0.6\linewidth]{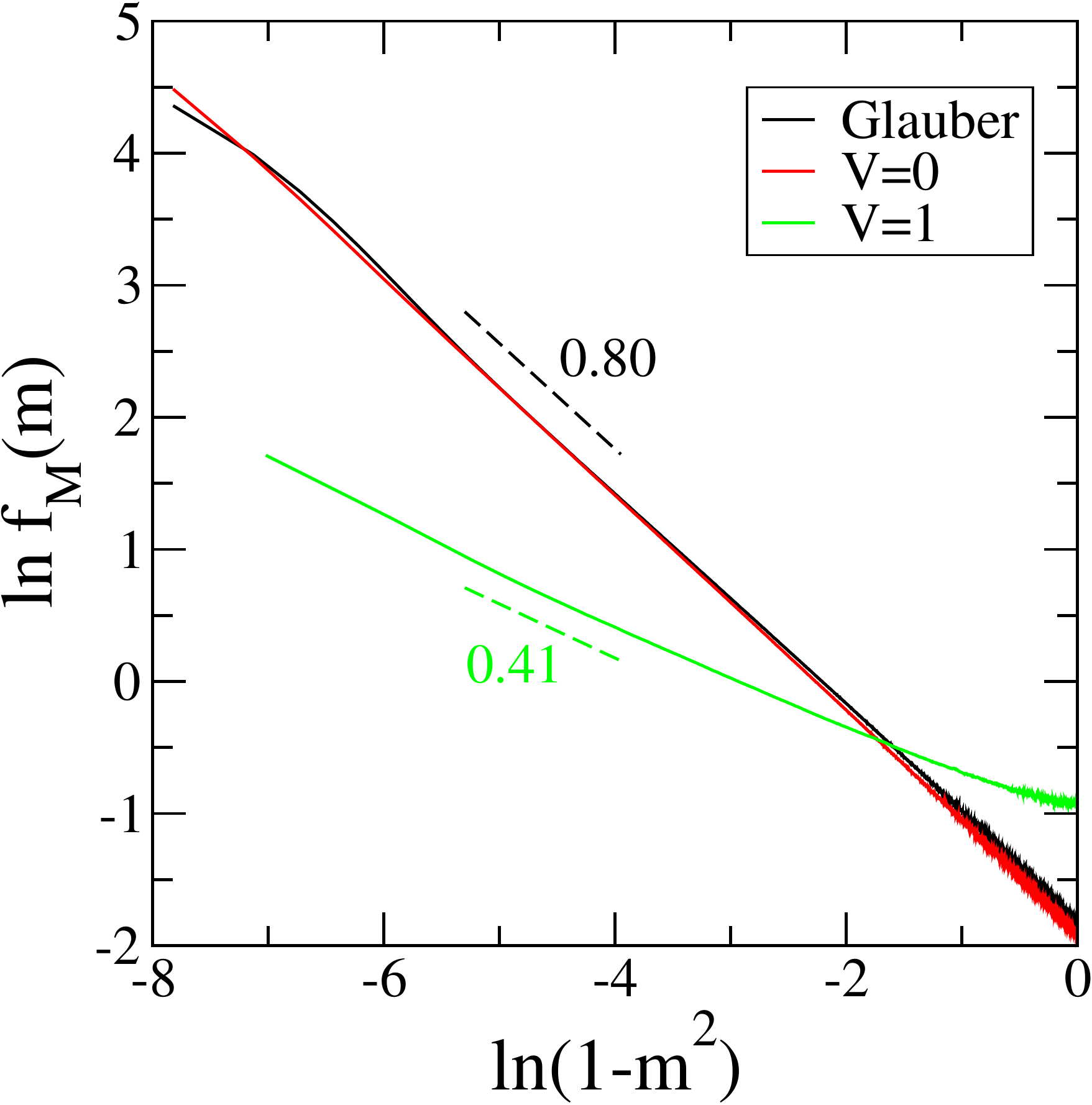}
\end{center}
\caption{Probability density $f_{M}(m)$ against $(1-m^2)$. 
The absolute values of the slopes give estimates for $1-\theta$ where $\theta$ is the persistence exponent.
The system sizes are $L=4096$,
and the data result from at least 500 independent histories.
The times used are $t=10\,000$ ($\xi_C(t)\approx 48 $) for Glauber and $V=0$ dynamics and $t=4480$ ($\xi_C(t)\approx 400$ as obtained by extrapolation of the data of figure~\ref{fig:xiC}) for $V=1$ dynamics.
}
\label{fig:fM}
\end{figure}

We performed simulations for the three dynamics under study (Glauber, $V=0$ and $V=1$)
with the aim of complementing the study of the persistence probability.
Figure~\ref{fig:fM} depict our results.
It yields the predictions $\theta\approx 0.20$ for Glauber and $V=0$ dynamics, and $\theta\approx 0.59$ for $V=1$ dynamics.
These values are compatible with the estimates obtained from the power-law decay of the persistence probability $p(t)$ in figure~\ref{fig:pers2}.
Note that figure~\ref{fig:fM} is also in favour of a common value for the persistence exponent of Glauber and $V=0$ dynamics.

\section{Discussion}
\label{sec:discussion}

\begin{figure}[ht]
\begin{center}
\includegraphics[angle=0,width=0.8\linewidth]{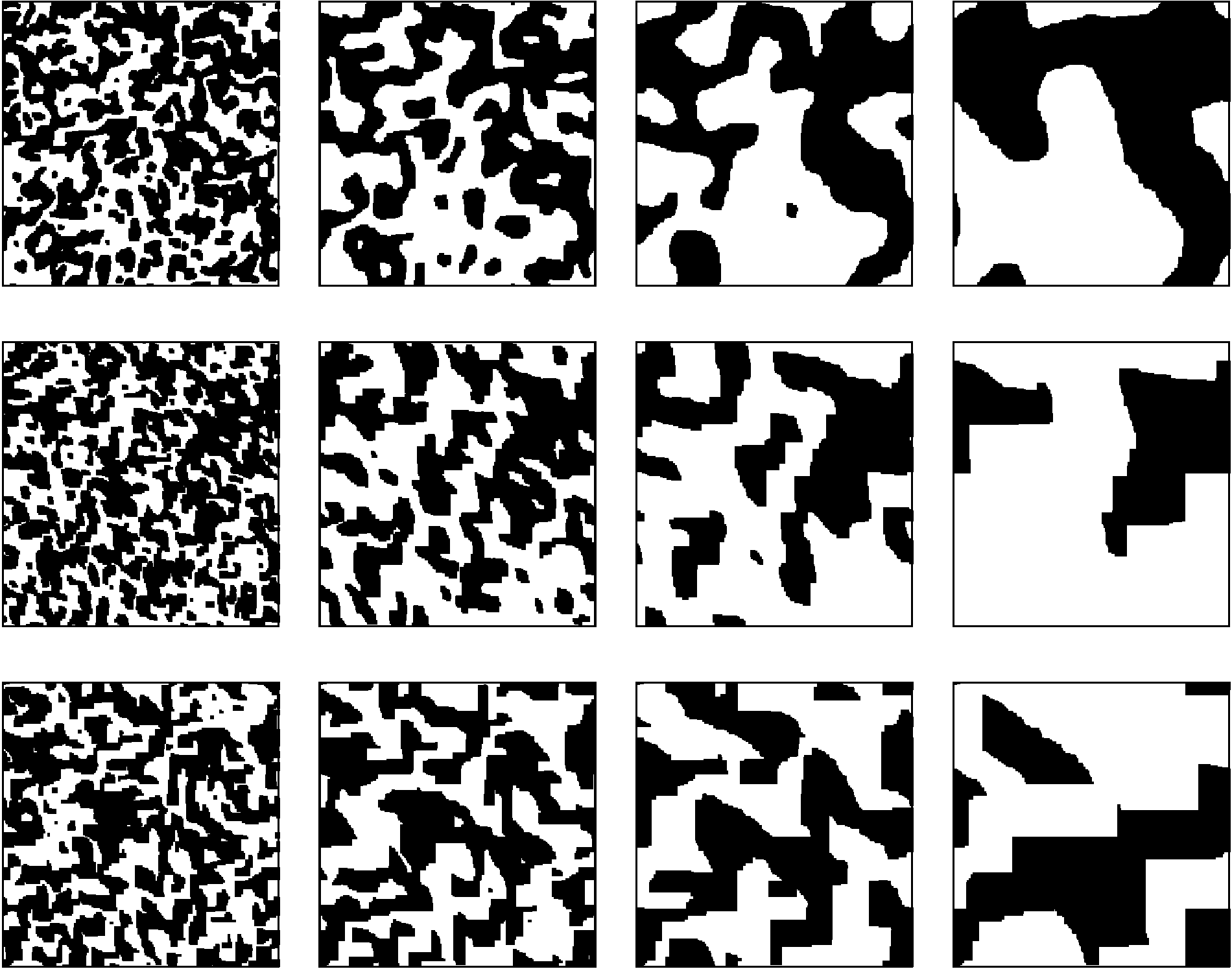}
\end{center}
\caption{Snapshots of a system of linear size $L=300$ for the three dynamics, Glauber (isotropic coarsening), $V=0$ (anisotropic coarsening), $V=1$ (ballistic coarsening).
Times are chosen such that, from left to right, $\xi_{C}(t)=4,8,16,32$, corresponding to $t=70, 280, 1120, 4680$ for the first two cases and $t=39, 78, 156, 312$ for the last one.
}
\label{fig:snapcoars}
\end{figure}

In the present work we have investigated the behaviour of an Ising spin system  on the two-dimensional square lattice undergoing phase ordering at zero temperature after a quench from infinite temperature.
This system of spins obeys dynamical rules defined by the rate function~(\ref{eq:rate}),
where the parameter $V$ controls the directedness or asymmetry of the influence of the neighbours on the flipping spin.
For $V=0$ the dynamics is reversible but not isotropic.
As soon as $V\ne0$ the dynamics becomes irreversible.
In the two special cases $V=\pm1$ the only influential spins on the central spin are, respectively, the North and East spins ($V=1$) or the South and West spins ($V=-1$).

The most salient outcome of the present study is the influence of irreversibility on phase ordering, which manifests itself by a crossover from diffusive coarsening to ballistic coarsening as soon as $V\ne0$.
Note that this change of regime, from diffusive to ballistic coarsening, does not exist either 
for the Ising chain~\cite{cg2011} or for the spherical model~\cite{spheric}, which only experience diffusive coarsening.

This change of regime was demonstrated by several means, namely
\begin{enumerate}
\item by the melting of a minority cluster,
\item by the behaviour of the growing length $\xi_{E}(t)$,
\item by the measurement of the times to reach the ground state or a blocked configuration,
\item by the behaviour of the growing length $\xi_{C}(t)$,
\item by the measurement of the autocorrelation and persistence exponents,
\item and by the investigation of the statistics of the mean magnetization.
\end{enumerate}

So doing we discovered that the self-similarity of the coarsening process, which holds for the usual reversible single-spin flip dynamics, still holds in the presence of irreversibility, as testified by the scaling of the equal-time correlation function and by the existence of non-trivial autocorrelation or persistence exponents.
The well-known patterns of coarsening for the usual Glauber dynamics are changed in the presence of an anisotropy in the dynamics.
This is illustrated in figure~\ref{fig:snapcoars}, which depicts the three situations studied in the present work: isotropic diffusive coarsening (Glauber dynamics), anisotropic diffusive coarsening ($V=0$), and ballistic coarsening ($V=1$).

The presence of metastable or blocked configurations investigated in~\cite{kr1,kr2,kr3,kr4,picco2} is also observed in the present model.
We plan to come back to this issue in the future.

Let us finally mention that the zero-temperature dynamics of the fully asymmetric model with $V=1$ is similar to that of the Toom model~\cite{toom} with zero noise and with random sequential dynamics.
In this context, the ballistic disappearance of a right triangle was noted in~\cite{grinstein1,grinstein2}.
The property of generic nonergodicity (i.e., two-phase coexistence over a finite region of the phase diagram) observed in the Toom model~\cite{toom,grinstein1,grinstein2} is thus expected to hold as well for the directed Ising model considered in the present work (see also~\cite{hinrich,munoz}).

\ack
It is a pleasure to thank T Bodineau, J M Luck and D Mukamel for helpful discussions.
MP acknowledges financial support by the US
National Science Foundation through grant DMR-1205309.

\end{document}